\documentstyle[aps,preprint,amsfonts]{revtex}
\renewcommand{\vec}[1]{\mbox{\boldmath$#1$}}

\newcommand{\be}{\begin{equation}}
\newcommand{\ee}{\end{equation}}
\newcommand{\bea}{\begin{eqnarray}}
\newcommand{\eea}{\end{eqnarray}}
\begin{document}
\draft
\title{SO(3) nonlinear {\boldmath $\sigma$} model for a doped 
quantum helimagnet}
\author{S.~Klee$^{1,2}$, A.~Muramatsu$^{2}$}
\address{$^{1}$Institut f\"ur Theoretische Physik, 
Universit\"at W\"urzburg, 97074 W\"urzburg, Germany}
\address{$^{2}$Institut f\"ur Physik, Universit\"at Augsburg, 
86135 Augsburg, Germany}
\maketitle
\begin{abstract}
A field theory describing the low-energy, long-wavelength sector of an 
incommensurate, spiral magnetic phase is derived from a spin-fermion model
that is commonly used as a microscopic model for high-temperature 
superconductors. After integrating out the fermions in a path-integral 
representation, a gradient expansion of the fermionic determinant is 
performed. This leads to an O(3)$\otimes$O(2)-symmetric quantum nonlinear 
$\sigma$ model, where the doping dependence 
is explicitly given by generalized fermionic susceptibilities which enter 
into the coupling constants of the $\sigma$ model and contain the fermionic
band-structure that results from the spiral background. A stability condition 
of the field theory self-consistently determines the spiral wavevector as a 
function of the doping concentration. Furthermore, terms of topological nature 
like the $\theta$-vacuum term in (1+1)-dimensional nonlinear $\sigma$ models 
are obtained for the plane of the spiral. 
\end{abstract}


\section{Introduction}
The properties of doped magnetic systems have been intensively studied in 
the context of strongly correlated electronic systems like heavy-fermion 
compounds and cuprate superconductors. The coexistence of local magnetic 
moments and itinerant fermions in these materials has raised many questions 
concerning the interplay between magnetic and charge degrees of freedom close 
to a magnetic instability. A powerful tool for a quantitative investigation of 
the critical behaviour of given micro\-scopic models is provided by 
renormalization group studies of the corresponding continuum theories. 

The field-theoretical approach has led to a successful description of the 
two-dimensional (2D) Heisenberg quantum antiferromagnet, where even a 
quantitative comparison with experimental results for the undoped parent 
compounds of the cuprates was achieved \cite{chn}. Results ob\-tained from 
studies of the appropriate continuum theory, the O(3) (2+1)-dimensional 
quantum nonlinear $\sigma$ model, through renormalization group analysis 
\cite{chn} and chiral perturbation theory \cite{HN}, agree very well with 
experimental \cite{End} and numerical \cite{ding} data.

More recently, field theories for frustrated 2D Heisenberg quantum 
antiferromagnets like the one on a triangular lattice 
\cite{domb,antimotri,apel,ADM93} were derived. 
Since the starting point is the corresponding N\'eel state with
a noncollinear order, the field theory is a (2+1)-dimensional SO(3) quantum 
nonlinear $\sigma$ model, as is generally expected for spiral states. 
Renormalization group analyses were performed for this model 
\cite{apel,azaria1} as well as for the classical O(3)$\otimes$O(2) 
nonlinear $\sigma$ model in $(2+\epsilon)$ dimensions \cite{azaria0,azaria2}. 
Furthermore, the scaling properties of the quantum phase 
transition from the ordered helical to the disordered phase were 
described \cite{chub}, based on scaling arguments and on a scenario of 
deconfined spinons. An understanding of the critical properties of 
frustrated spin systems between two and four dimensions was also achieved 
recently \cite{azalm}. Thus, a fairly large amount of theoretical predictions 
was obtained in recent years in the case of pure quantum spin systems. 

However, the most interesting and relevant situation of strongly correlated 
systems, where magnetic as well as charge degrees of freedom interact, was
until now investigated to a much lesser extent. Apart from numerous 
mean-field attempts to analyze those systems, to the best knowledge of 
the authors only few field-theories have 
been derived from microscopic models so far \cite{schulz,muze,mulett,chris}, 
besides phenomenological approaches 
\cite{shankar}, where fluctuations effects are duly taken into account. 

The studies above were performed for holes in an antiferromagnetic background, 
such that no equivalent descriptions are available yet for the region of the 
phase diagram of strongly correlated systems, like some of the cuprates, in 
which incommensurate spin fluctuations appear.
For doping concentrations in the superconducting regime, neutron scattering 
experiments have shown \cite{Cheo,thur92} that the Lanthanum compounds 
exhibit peaks in the magnetic scattering intensity at wave vectors which are 
shifted by $\pm\delta (\pi,0)$ and $\pm\delta (0,\pi)$ from the 
antiferromagnetic point, where the shift $\delta$ increases with the doping.
The correlation length of the incommensurate fluctuations has been found to 
be considerably larger \cite{thur92} than the lattice spacing, which makes a 
continuum approach reasonable. Our aim here is to establish a mathematically 
rigorous connection between a suitable, sufficiently general microscopic 
model and a continuum theory. Considering fluctuations around a helical spin 
configuration, we will develop a field theory that, in conjunction with the 
above-mentioned recent progress in renormalization group studies, leads to a   
quantitative description of the critical behaviour of an incommensurate spin 
system interacting with doped charge carriers.

Previous theoretical studies of microscopic models of strongly correlated 
fermions suggest that the competition of hopping and exchange effects may 
lead to the formation of a spiral phase in the spin subsystem. Mean-field 
calculations favour spiral against antiferromagnetic configurations in Hubbard
models for small but non-zero doping \cite{schulz,MF}. 
Concentrating on the one-hole 
case, Shraiman and Siggia proposed spiral spin ordering based on a 
phenomenological continuum description of the $t$-$J$ model 
\cite{shraimansiggia}. They argued that the introduction of a low density of 
doped holes into a locally antiferromagnetic background may lead to an 
incommensurate, helical rotation of the staggered magnetization as a 
consequence of a coupling between the spin current of the holes and the 
magnetization current of the background. Such a coupling was in fact derived
in a continuum action obtained directly from the spin-fermion 
model \cite{chris}. Later on, the phase diagram of the Shraiman-Siggia model 
was investigated within a $\frac{1}{N}$ expansion \cite{sachd}. A transition 
from the commensurate to an  incommensurate magnetic phase was obtained with 
increasing doping, both in the ordered and in the quantum-disordered regime
for suitable values of the Shraiman-Siggia coupling. However, apart from the 
fact that the starting model is a phenomenological one, it is not clear to us 
to which extent those results really apply to the case of a spiral phase, 
since the conclusions on the critical behavior are derived on the basis of 
an order-parameter in the manifold $S^2$ and not SO(3), as it should be in
the non-collinear case. Finally, it should be mentioned
that incommensurate spin configurations have further been 
shown to be induced by doping in numerical simulations of the one-band and 
the three-band Hubbard model away from half-filling \cite{QMCmor,QMCdo,QMCfa}.

The starting point of our work is the spin-fermion Hamiltonian \cite{sf}, 
which describes mobile fermions interacting with a background of localized 
spins through an exchange term. It is a generalization of the Kondo 
lattice, since also a Heisenberg exchange interaction among the localized 
spins is included. We will concentrate on the two dimensional case that 
gives a realistic description of the cuprates. A central aspect that 
directs the choice towards the spin-fermion model is its being analytically 
tractable, an important advantage over other microscopic models for strongly 
correlated fermions. The study follows essentially the same steps as in the
antiferromagnetic case \cite{muze,mulett}, although the actual calculation
(Sec.\ IV) greatly differs from that case. In a path-integral representation 
of the model, the fermionic degrees of freedom appear bilinearly and are 
integrated out exactly. An action containing only the spin degrees of 
freedom is obtained in terms of a fermion determinant and a pure spin part. We 
consider incommensurate, short-ranged spiral configurations for the spins and 
discuss their parametrization for long-wavelength, low-energy fluctuations 
around the ordered spiral phase. The gradient expansion of the fermion 
determinant is carried out in energy and momentum space in 
such a way that the occurring infinite series can be summed to all orders of 
the coupling constant by using the constraint on the order parameter. We show 
that in the limit of long wavelengths and low energies of the spins, the 
continuum theory is not only given by an SO(3) quantum nonlinear $\sigma$ 
model, as assumed in previous phenomenological approaches, but additional 
terms appear. On the one hand, a term linear in derivatives is obtained, 
such that parity is explicitly broken. However, since this term is not 
positive definite, it should vanish in order to guarantee a stable spiral 
configuration. This requirement leads to two equations
that determine the wavevector for the spiral as a function of doping and 
the parameters of the model. On the other hand, a term of topological 
character like the one obtained in the continuum limit of the 
one-dimensional antiferromagnetic Heisenberg model \cite{theta} appears. 
However, in contrast to that case, the corresponding coupling constant is 
doping dependent, and hence, it can vary continuously. Therefore, our gradient 
expansion not only delivers the coupling constants of the SO(3) quantum 
nonlinear $\sigma$ model as functions of the microscopic parameters and 
generalized fermionic susceptibilities which contain the doping dependence,
but shows important differences with phenomenological models obtained only 
on the basis of symmetry arguments. Finally, we would like to add that the 
present approach does not impose any restrictions on the energy and 
momentum scales of the fermions.  
\section{Microscopic model and path-integral description}
We consider a spin-fermion Hamiltonian which describes spins
localized on the vertices of a square lattice ("Cu-sites") interacting 
with fermions moving between sites situated on the bonds ("O-sites").
The problem remains nontrivial due to the coupling of the band
fermions to the surrounding localized spins. The Hamiltonian is
\begin{eqnarray}
H_{\rm sf}&=& \sum_{\stackrel{<jj'>}{\alpha}} t_{jj'}\,\,
c_{j \alpha}^{\dagger}c_{j' \alpha}
+\sum_{i} \,\Bigl( 
\sum_{\stackrel{<jj',i>}{\alpha,\alpha'}}
J_{\rm K}^{jj',i}\,\,c_{j \alpha}^{\dagger}\,
\vec{\sigma}_{\alpha \alpha'}c_{j' \alpha'}\Bigr) 
\cdot\vec{S_i} \nonumber \\ 
& &+ \;J_{\rm H}\sum_{<ii'>} \vec{S_i}\cdot\vec{S_{i'}}\;.
 \end{eqnarray}
Here $c_{j,\alpha}^{\dagger}$ and $c_{j,\alpha}$ are creation and
annihilation operators, respectively, for holes with spin 
projection $\alpha=\uparrow,\downarrow$ on 
sites situated on the bonds of the square lattice, denoted by 
the index $j$.
The index $i$ runs over the vertices of the  square lattice. 
The kinetic term  describes hopping processes between the sites
situated on the bonds, and it can contain a direct hopping  
between sites $j$ and $j'$ as well as an an effective
hopping between  sites $j$ and $j'$ mediated by a central
site $i$. 
The term proportional to $J_{\rm K}$ is a non-local, Kondo-like
interaction between the localized spins $\vec{S_i}$
and the holes on the neighbouring sites. It includes
spin-exchange processes with hopping in addition to
pure exchange processes.
The vector $\vec{\sigma}_{\alpha \alpha'}$
consists of the three
Pauli matrices. The summation denoted by $<jj',i>$ runs over all
pairs of  sites $j$, $j'$ which are nearest neighbours to a 
given site $i$.
Finally, $H_{\rm sf}$  contains an antiferromagnetic Heisenberg
superexchange interaction between nearest-neighbour spins
$\vec{S_i}$.

The spin-fermion model can be obtained as the strong-coupling 
limit \cite{sf} of the three-band Hubbard model \cite{Eme}, which,
according to numerical simulations \cite{QMCdo,dopf}, consistently describes 
the CuO$_2$ planes of the cuprates. Limiting cases of the spin-fermion 
model correspond to other models that are frequently discussed in the 
context of strongly correlated fermion systems. In the limit of zero doping, 
the spin-fermion model reduces to an antiferromagnetic S=$\frac{1}{2}$ 
Heisenberg model, which yields a quantitative description of the undoped
cuprates \cite{chn,HN,ding}. For large $J_{\rm K}$, the doped holes on the
O sites strongly bind to the central Cu ion to form a local singlet, so that 
for zero direct O-O hopping the low-energy dynamics of $H_{\rm sf}$ can be 
mapped onto the $t$-$J$ model \cite{zhari}. For finite O-O hopping, the 
spin-fermion Hamiltonian can be mapped onto a generalized $t$-$J$ model 
containing second and third nearest-neighbour hopping and spin-flip
hopping \cite{aligia}. For vanishing Heisenberg interaction, 
$H_{\rm sf}$ is equivalent to a Kondo-lattice Hamiltonian with a nonlocal
exchange between band fermions and localized spins.

In the following we discuss only briefly the path-integral representation 
of the model in order to set up the notation, since the same steps were already 
performed in the antiferromagnetic case \cite{muze}. We represent the partition 
function of the spin-fermion Hamiltonian as an imaginary-time path-integral.
Employing a spin-$\frac{1}{2}$ coherent state representation \cite{klau} for 
the spin degrees of freedom and Grassmann variables for the hole degrees of 
freedom, we obtain for the partition function
of $H_{\rm sf}$
\begin{equation}
Z_{\rm sf}= \!\int\limits_{{\scriptsize \vec{S}
(0)=\vec{S}(\beta)}}\!\!\!\! {\cal
D}\vec{S} \int\limits_{c(0)=-c(\beta)}\!\!\!\!\! {\cal
D}c^* {\cal D}c\,\,\,  \exp \left( S_{\rm s} + S_{\rm f}\right)
\,\,, \label{Zsf1}
\end{equation}
where $\beta= 1/k_BT$.  
In Eq.~(\ref{Zsf1}), the pure spin part of the action is given by
\begin{equation}
S_{\rm s} =  \int_0^{\beta}{\rm d}\tau \left[i \sum_i
\vec{A}\left( \vec{S}_i (\tau)/S\right) \cdot
\partial_{\tau}\vec{S}_i(\tau)
-J_{\rm H}\sum_{<ii'>} 
\vec{S}_i(\tau) \cdot \vec{S}_{i'}(\tau) \right] \;,
\label{Sspin}
 \end{equation}
where now $\vec{S}=S \vec{\Omega}$ and $\vec{\Omega}=
(\sin \theta_i(\tau) \cos \phi_i(\tau), \sin
\theta_i(\tau) \sin\phi_i(\tau),  \cos \theta_i(\tau))$. 
As is well known, the 
first term in the action  $S_{\rm s}$ is the Berry phase 
for the adiabatic transport of a quantum spin along a closed circuit
and responsible for the correct 
quantization of the spins. The monopole potential $\vec{A}$ 
satisfies the constraint
\begin{equation}
\epsilon^{abc}\frac{\partial A^{b}}{\partial {\Omega}^c}={\Omega}^a 
\label{monpolpot}
\end{equation}
on the unit sphere. The measure for the integration over the
spin variables is given by ${\cal
D}\vec{S}=\prod_{i,\tau}\frac{(2S+1)}{4\pi}\,
\sin\theta_i(\tau)\,d\theta_i(\tau)\,d\phi_i(\tau)$.

In momentum and frequency space, and after diagonalizing
the kinetic part of $H_{\rm sf}$, which yields two bands 
with a dispersion relation $\epsilon(\vec{k},\lambda)$,
$\lambda=1,2$, for which we refer to Ref.~\cite{muze}, we obtain
\begin{eqnarray}
S_{\rm f}&=&  \beta \sum_{\alpha,\alpha'}
\sum_{k,k'} \sum_{\lambda,\lambda'} c_{\alpha k \lambda}^*
\biggl[ \Bigl(i\epsilon_n-\epsilon(\vec{k},\lambda)\Bigr)\delta_{\alpha
\alpha'}\, \delta_{k k'}\,\delta_{\lambda \lambda'} 
\nonumber \\
& &-{}4J_K\vec{\sigma}_{\alpha
\alpha'} \cdot
\vec{S}_{k-k'}s^*(\vec{k},\lambda)\,s(\vec{k}',\lambda')\biggr]
c_{\alpha' k' \lambda'} \;, \label{Sfer}
\end{eqnarray}
where $k=(\epsilon_n,\vec{k})$. Its zeroth component is a discrete
Matsubara frequency defined by $\epsilon_n=(2n-1)\pi/\beta$ for fermionic 
fields, and by $\omega_n=2n\pi/\beta$ for bosonic fields.
The form factors $s(\vec k,\lambda)$ contain information on
the band structure of the free system and are given by 
$s(\vec k,\lambda)= \mbox{e}_1({\vec{k}, \lambda}) \,\sin \left(
k_x a/2 \right) + \mbox{e}_2 ({\vec{k}, \lambda}) \,\sin \left(
k_y a/2 \right)$, where  $\mbox{e}_{1,2}({\vec{k}, \lambda})$ 
are the components of the eigenvectors\cite{muze} that diagonalize
the kinetic part of $H_{\rm sf}$.

Since the action in Eq.~(\ref{Sfer}) is bilinear in the Grassmann
fields, the integration ${\cal D}c^*{\cal D}c$ 
may be carried out within the path-integral to obtain 
\begin{equation}
Z_{\rm sf}=\!\int\limits_{{\scriptsize \vec{S}(0)=\vec{S}(\beta)}}
\!\!\!\! 
{\cal D}\vec{S}  \exp \left( S_{\rm s} +\mbox{Tr}\ln G^{-1}\right)\;,
\label{Zsf}
\end{equation}
where
\begin{equation}
G^{-1}=G_0^{-1}-\Sigma \label{Ginv}
\end{equation}
is the inverse propagator of the fermions in the presence 
of the dynamical spin field $\vec{S}_{k-k'}$.
The free fermionic propagator is given by 
\begin{equation}
\bigl(G_0^{-1}\bigr)_{\alpha k \lambda, \alpha' k'
\lambda'}=\bigl(i\epsilon_n-\epsilon(\vec{k},\lambda)\bigr)\,
\delta_{\alpha \alpha'}
\delta_{k k'}\delta_{\lambda \lambda'}\;,
\end{equation}
and $\Sigma$ is the self-energy 
of the holes interacting with the Cu spins,
\begin{equation}
\bigl( \Sigma \bigr)_{\alpha k \lambda, \alpha' k'
\lambda'}=4J_K\, \vec{\sigma}_{\alpha \alpha'}\vec{S}_{k-k'}\,
s^{\dagger} (\vec{k},\lambda)s(\vec{k}',\lambda')\;.
\label{selfen}
\end{equation}
The trace in Eq.~(\ref{Zsf}) is to be taken over the indices
$k$, $\alpha$ and $\lambda$. It is convenient to shift 
the form factors $s(\vec{k},\lambda)$ from the self-energy 
$\Sigma$ to 
the unperturbed fermionic propagator $G_0$. 
It can be checked that 
by writing the logarithm as a power series, 
$\mbox{Tr} \ln \bigl(G_0^{-1}-\Sigma \bigr)= \mbox{Tr} \ln
\bigl(G_0^{-1}\bigr) - \sum_{n=1}^{\infty} \frac{1}{n}
\mbox{Tr}\bigl(G_0\Sigma\bigr)^n$, and rearranging the terms in
the matrix product,
a redefinition
of the free fermionic propagators 
\begin{equation}
\bigl(\bar{G}_0\bigr)_{\alpha k, \alpha' k'}=\sum_{\lambda=1}^2
\frac{|s(\vec{k},\lambda)|^2}{\bigl(i\epsilon_n-\epsilon
(\vec{k},\lambda)\bigr)}\, \delta_{\alpha \alpha'}\delta_{k k'}
\;, \label{G0}
\end{equation}
and the self-energy
\begin{equation}
\bigl(\bar{\Sigma} \bigr)_{\alpha k, \alpha' k'}
=4J_K\,\vec{\sigma}_{\alpha
\alpha'} \cdot \vec{S}_{k-k'}\;, \label{Sigma}
\end{equation}
yields the relation
\begin{equation}
\mbox{\raisebox{-8pt}{$\stackrel{\textstyle \mbox{Tr}}{\scriptstyle
(\alpha,k,\lambda)}$}}\ln \bigl(G_0^{-1}-\Sigma\bigr)=
\mbox{\raisebox{-8pt}{$\stackrel{\textstyle
\mbox{Tr}}{\scriptstyle(\alpha,k)}$}} \ln \Bigl(\sum_{\lambda=1}^2
|s(\vec{k},\lambda)|^2 G_0^{-1}(3-\lambda) \Bigr)
+\mbox{\raisebox{-8pt}{$\stackrel{\textstyle
\mbox{Tr}}{\scriptstyle(\alpha,k)}$}} \ln
\bigl(\bar{G}_0^{-1}-\bar{\Sigma} \bigr)\;. \label{umdef}
\end{equation}
Since the first term on the right-hand side of Eq.~(\ref{umdef})
 contains only the
free propagator and no interaction contribution, it  may be regarded
as a normalization constant and 
will be ignored from now on. Note that the trace in
the second term on the right-hand side of Eq.~(\ref{umdef})
no longer includes the band index $\lambda$. Since in the
following we will use the free fermionic propagator and the 
fermionic self-energy only in the forms (\ref{G0}) and
(\ref{Sigma}), we will omit the bars to simplify the notation,
and refer to
Eqs.~(\ref{G0}) and (\ref{Sigma}) simply as $G_0$ and $\Sigma$.
We further introduce the abbreviation
\be
g_0(k)=\sum_{\lambda=1}^2
\frac{|s(\vec{k},\lambda)|^2}{\bigl(i\epsilon_n-\epsilon
(\vec{k},\lambda)\bigr)}\,.
\ee
Later on we will need the momentum-space symmetry properties 
of the function
$g_0(k)=g_0(\epsilon_n,\vec{k})$. Using
the explicit expressions \cite{muze} for the energy eigenvalues and 
eigenvectors of the free fermion system,
the following symmetry relations can be readily deduced:
\bea
g_0(\epsilon_n,\vec{k})& =&g_0(\epsilon_n,-\vec{k})\;,\label{g0sym1} \\
g_0(\epsilon_n,k_x,k_y)&=&g_0(\epsilon_n,-k_x,k_y)\;\,=\;\,
g_0(\epsilon_n,k_x,-k_y)\label{g0sym2}\;.
\eea
We have derived an action where the only degrees of freedom 
appearing explicitly are the spins, which are suitable for a continuum
approximation. This continuum approximation does
not affect the fermionic degrees of freedom which are taken into
account with their full dispersion relation. 
\section{Order parameter for the spiral}
For the spin configuration $\vec{S}_i$ along the lattice, we
have to introduce an expression characterizing
the spiral order expected in the classical
ground state. We can then take into account
long-wavelength, low-energy fluctuations around this ordered
helical state. In the classical ground state, the spins lie in a
plane in spin space. In order to derive a continuum theory, we need to
identify vector fields which are smooth in the long-wavelength,
low-energy regime. In the case of a Heisenberg antiferromagnet on a triangular 
lattice, the constituent fields of the order parameter can be obtained
from linear combinations of the physical spins with\-in one magnetic cell 
\cite{domb,apel}, in analogy to the construction of the order parameter fields 
describing collinear antiferromagnet configurations \cite{hald83}.
In the case of an incommensurate helix however, there are
in\-fi\-nitely many sublattices and a magnetic cell does not exist.

In order to construct the order parameter for the spiral, we consider 
long-wavelength fluctuations around a classical spiral configuration with 
wave-vector
\be
\vec{k}_s=\vec{Q}+\delta\vec{Q} \label{ks} \; ,
\ee
where
\be
\vec{Q}=\left( \frac{\pi}{a},\frac{\pi}{a} \right) \label{ku}
\ee
is the wave vector of
the antiferromagnetic state, while 
\be
\delta\vec{Q}=(\delta 
Q_x,\delta Q_y) \label{dku}
\ee
is incommensurate with the lattice, i.e. $m \,\delta\vec{Q} \neq n \,\vec{Q}$ 
for all $m$, $n$ integer. We start with the vector product of neighbouring 
spins. We define first a vector 
$\vec{n}_{3i} = \vec{S}_i \times \vec{S}_{i+1}\,/N_i$,  where
$N_i=S^2\, \sin \phi_{i,i+1}$ and $\phi_{i,i+1}$ is the angle enclosed 
between the spins $\vec{S}_i$ and $\vec{S}_{i+1}$. This choice is well defined
as long as the line joining the two spins is not perpendicular to $\vec{k}_s$.
The vector $\vec{n}_3$ reduces in the classical ground state to a vector 
parallel to the axis of the helix, which is independent of the lattice site. 
We now define a matrix 
$R^{ab}_i= \cos (\vec{k}_s \! \cdot \!\vec{r}_i) \, \delta^{ab} + 
[1- \cos (\vec{k}_s \! \cdot \!\vec{r}_i)]\, n_{3i}^a n_{3i}^b 
- \sin (\vec{k}_s \! \cdot \!\vec{r}_i) \, \epsilon^{abc} \, n_{3i}^c $ 
which performs rotations about $\vec{n}_{3i}$. Here the upper indices 
$a,b,c=1,2,3$ denote components in spin space. 
We adopt the convention that summation over repeated indices is implied, 
although for clarity we will occasionally write the summation sign explicitly.
Applying $\hat{R}_i$ to the spin $\vec{S}_i$ rotates it back into a 
space-fixed axis, which provides the definition for a second vector being 
constant in the classical ground state: 
$\vec{n}_{1i}= \hat{R}_i\,\vec{S}_i\,/S$. Defining a third vector 
$\vec{n}_{2i}= \vec{n}_{3i} \times \vec{n}_{1i} \label{n123}$, 
we have a complete set of orthonormal vectors which are constant in the 
classical ground state and fulfill 
\begin{equation}
n^a_b \, n^a_c = \delta_{bc} \label{constr}\;.
\end{equation}
Using the above basis, we parametrize the incommensurate spiral configuration 
as
\begin{equation}
\vec{S}_{pq} (\tau) =S\,\left[ \vec{n}_1 (p,q,\tau) \cos\, (\vec{k}_s \! 
\cdot \!\vec{r}_{pq})-\vec{n}_2 (p,q,\tau) \sin\, (\vec{k}_s \!\cdot \!
\vec{r}_{pq})\right] \;, \label{GS}
\end{equation}
where now $p$, $q$ (instead of $i$) denote the sites on the two-dimensional
square lattice and $\vec{r}_{pq}=(p\cdot a,q\cdot a)$, with $a$ the lattice 
constant. It is evident from Eq.~(\ref{GS}) that in the limit 
$\delta\vec{Q} \rightarrow 0$, the antiferromagnetic ground state is 
reproduced, so that $\delta\vec{Q}$ describes the deviation from perfect 
N\'{e}el order.

In order to evaluate the continuum limit of the action, we allowed 
the vectors  $\vec{n}_1$ and $\vec{n}_2$ to be smooth functions
of the lattice site and of the Euclidean time $\tau$. The constraint 
(\ref{constr}) must be satisfied at every given point in space-time. 
In a previous work \cite{dipapp}, we have treated the special case of planar 
fluctuations, where the variation of $\vec n_1(p,q,\tau)$ and 
$\vec n_2(p,q,\tau)$ was confined to the plane  which is defined by 
$\vec n_1$ and $\vec n_2$ in the classical ground state. The order parameter 
for the description of planar fluctuations around a spiral configuration is a 
two-component unit vector in the manifold $S_1$. The corresponding field 
theory was shown to be an  O(2) nonlinear $\sigma$ model. Here we extend this 
treatment to the general case of fully three-dimensional fluctuations. This
generalization turns out to be highly non-trivial (see Sec.\ IV and V). In 
the present case, fluctuations of $\vec n_1(p,q,\tau)$ and 
$\vec n_2(p,q,\tau)$ imply fluctuations of $\vec n_3(p,q,\tau)$, so that the 
excitations are described by an SO(3) order parameter:
\begin{equation}
\hat{Q} = \left( \vec n_1 \, \vec n_2\, \vec n_3 \right)
\mbox{\hspace*{1cm}}\mbox{or}\mbox{\hspace*{1cm}}
Q_{ab}=n^a_b\,. \label{SO3}
\end{equation}
In contrast to an antiferromagnetic configuration,
where the ground state is
invariant under rotations about the spin axis, in a
noncollinear  ground state global rotations about any axis lead
to a new ground state. Thus in the spiral phase
the global O(3) rotational symmetry in spin
space is completely broken and we expect three Goldstone modes.

In helical phases, however, there is still a {\em local} rotational symmetry, 
which was first identified in Ref.~\cite{antimo_1D_dD}: $\vec{S}_{pq}(\tau)$ 
is invariant under rotations by an arbitrary local angle $\psi_{pq}(\tau) $
about the {\em local} axis $\vec{S}_{pq}(\tau)$. This can be readily seen in 
a Schwinger boson representation of the spins 
$\vec{S}=S\, \bar{\omega}^{\alpha} \vec{\sigma}^{\alpha \beta} \omega^{\beta}$, 
where $\omega^{\alpha}$ is a doublet of complex scalar fields satisfying 
$\bar{\omega}^{\alpha} \omega^{\alpha}=1$ \cite{antimo_1D_dD}. In this 
representation, a local rotation about the local spin axis corresponds to 
the U(1) gauge transformation $\omega \rightarrow e^{i\psi/2} \omega$. 
However, under these gauge transformations the order parameter fields 
$\vec{n}_i$ ($i = 1,2,3$)
are not invariant. In the general case, in which $\delta \vec{Q}$ 
is finite, an infinitesimal change in the angle $\psi_{pq}(\tau)$ between two 
sites will lead to a finite change in the fields $\vec{n}_i$. If the fields 
$\vec{n}_i$ are slowly varying in one choice of gauge, they will have rapid 
variations in other gauges. Thus, by focusing on continuous configurations,
one explicitly breaks this local symmetry by fixing the gauge. Our prescription
to determine the fields $\vec{n}_i$, and therefore to fix the gauge is 
physically natural, since it is given by the actual configurations 
of the spin fields $\vec{S}$. 
\section{Gradient expansion for the doped system}
In order to
parametrize the low-lying modes around the ground state of the
classical spiral, we decompose the spin field into a helical
and a uniform component so that
\begin{equation}
\vec{S}=S\,\left(\vec{n}+a\vec{L}\right)
\left(1+2\,a\,\vec{n}\cdot\vec{L}+a^2L^2\right)^{-1/2}\,.
\label{ansatz}
\end{equation}
where we have introduced the abbreviation
\begin{equation}
\vec{n}=\vec{n}_1 \cos (\vec{k}_s\cdot \vec{r}_{pq})-
\vec{n}_2\sin(\vec{k}_s\cdot\vec{r}_{pq})
\end{equation}
and the $(p,q,\tau)$-dependence of the fields is implicit.
$\vec{L}$ is a slowly varying ferromagnetic field with
$|a\vec{L}|\ll 1$, which corresponds to the net magnetization
density. $\vec{L}$ is of the same order as a first-order
derivative of $\vec{n}$. While $\vec{n}$ describes fluctuations around
the spiral wave vector $\vec{k}_s$, $\vec{L}$ describes
fluctuations around $\vec{k}=0$.
Since the resulting action will be bilinear in the ferromagnetic
fluctuation, $\vec{L}$ can be integrated out
at the end of the calculation.
The inverse square root factor in Eq.~(\ref{ansatz}) puts
$\vec{S}/S$ on the unit sphere. 

Expanding the ansatz for the spin field (\ref{ansatz})
up to second order in $a$ gives
\begin{equation}
\vec{S}= S\,\biggl\{ \vec{n}+ a\Bigl[ \vec{L}-
(\vec{n}\cdot\vec{L})\,\vec{n}\Bigr] 
-a^2\Bigl[ (\vec{n}\cdot\vec{L})\, \vec{L} 
+\frac{1}{2}\vec{L}^2\, \vec{n}
-\frac{3}{2}(\vec{n}\cdot\vec{L})^2\, \vec{n}
\Bigr]\biggr\}\,. \label{ansexp} 
\end{equation}
This expression will now be employed for the
space-time-dependent
spin field in the path-integral, Eq.~(\ref{Zsf}).
 We will perform the gradient
expansion of the pure spin part of the action,
Eq.~(\ref{Sspin}),
in real space. It is technically advantageous
to carry out the
gradient expansion of the fermion determinant
(see Eqs.~(\ref{Zsf})--(\ref{Ginv}))
in Fourier space, where it is approximately diagonal for
long-wavelength spin fields.
We  therefore transform the expression Eq.~(\ref{ansexp}) for
$\vec{S}_{pq}(\tau)$ into $k$-space (where
$k=(\epsilon_n,\vec{k})$) and insert it
into the fermionic self-energy $\Sigma$ (see Eq.~(\ref{Sigma})).
This leads to an expansion of the fermionic self-energy
in powers of $a$:
\begin{equation}
\Sigma=\Sigma^{(0)}+a\,\Sigma^{(1)}+a^2\,\Sigma^{(2)}\,.
\end{equation}
Introducing the abbreviations:
\begin{equation}
\vec{n}_{\pm}=\vec{n_1} \pm i\,\vec{n}_2 \,,
\end{equation}
we obtain for the zeroth, first and second orders of the
fermionic self-energy:
\begin{eqnarray}
\left( \Sigma^{(0)}\right)_{\alpha k, \alpha' k'}&=&\frac{g}{2}\,
\sigma^a_{\alpha \alpha'}\, \Bigl[\, n^a_-(k\!-\!k'\!-\!k_s)
+n^a_+(k\!-\!k'\!+\!k_s) \, \Bigr]\,, \label{Sigma0} \\
\left( \Sigma^{(1)}\right)_{\alpha k, \alpha' k'}&=&g\,
\sigma^a_{\alpha \alpha'}\,\Biggl\{ \,L^a(k\!-\!k')
-\frac{1}{4} \sum_{q_1,q_2} L^b(q_1) \times \,\nonumber \\
& &\biggl[  \,\,
\sum_{r=-,+} n_r^b(q_2\!+\!rk_s)\,
n_r^a(k\!-\!k'\!+\!rk_s\!-\!{\textstyle \sum_{i=1}^2}
q_i)\nonumber \\
& &{} +2 \sum_{d=1,2}\,n_d^b(q_2) \,
n_d^a(k\!-\!k'\! -\!{\textstyle \sum_{i=1}^2 q_i}) 
\,\biggr] \Biggr\}\,, \label{Sigma1} \\
\left( \Sigma^{(2)}\right)_{\alpha k, \alpha' k'}&=& 
g\,\sigma^a_{\alpha \alpha'}\,\Biggl\{  
-\frac{1}{2} \sum_{q_1,q_2}
L^a(q_1)\,L^b(q_2) \sum_{r=-,+} n_r^b(k\!-\!k'\!+\!rk_s\!
-\!{\textstyle \sum_{i=1}^2 q_i}) \nonumber \\
& &{}-\frac{1}{4} \sum_{q_1,q_2} L^b(q_1)
\,L^b(q_2) \sum_{r=-,+} n_r^a(k\!-\!k'\!+\!rk_s\!
-\!{\textstyle \sum_{i=1}^2 q_i})\nonumber \\
& &{}+\frac{3}{16} 
\sum_{q_1 \ldots q_4} L^b(q_1)\,L^c(q_2) \times\nonumber \\
& & \biggl[\;
\Bigl(\, \sum_{d=1}^2 n_d^b(q_3)
\,n_d^c(q_4) \Bigr) \sum_{r=-,+} n_r^a(k\!-\!k'\!+\!rk_s\!
-\!{\textstyle \sum_{i=1}^4 q_i})\nonumber \\
& &{}+2 \,\Bigl( \sum_{d=1}^2 n_d^a(q_3)
\,n_d^b(q_4) \Bigr) \sum_{r=-,+} n_r^c(k\!-\!k'\!+\!rk_s\!
-\!{\textstyle \sum_{i=1}^4 q_i}) \nonumber \\
& &+\sum_{r=-,+}
n_r^b(q_3\!+\!rk_s)\,n_r^c(q_4\!+\!rk_s)\,
n_r^a(k\!-\!k'\!+\!rk_s\!-\!{\textstyle
\sum_{i=1}^4q_i})\,
\biggl] \Biggl\}\,, \label{Sigma2} 
\end{eqnarray}
where we have defined the coupling constant
\begin{equation}
g=4J_K S\,.
\end{equation}
Note that $k_s=(0,\vec{k}_s)$.
Inserting this expansion of the fermionic self-energy into the
fermion determinant from Eq.~(\ref{Zsf}), it may be written as:
\begin{equation}
\mbox{Tr} \ln (G_0^{-1}-\Sigma) = 
\mbox{Tr} \ln (\tilde{G}_0^{-1})+\mbox{Tr} \ln
(1-a\tilde{G}_0\Sigma^{(1)}-a^2\tilde{G}_0\Sigma^{(2)})\,,
\label{ferdet}
\end{equation}
where
\begin{equation}
\tilde{G}_0^{-1}=G_0^{-1}-\Sigma^{(0)}\,. 
\end{equation}
The first term in Eq.~(\ref{ferdet}) represents the helical
contribution since the field $\vec{L}$ does not enter in it.
The second term contains the contributions of the ferromagnetic
field. 
The  helical and the ferromagnetic contributions to the
fermion determinant, Eq.~(\ref{ferdet}),
will be evaluated using different methods, which both
rely crucially on the spin-field momenta being small.
\subsection{Spiral contribution to the fermion determinant}
As a first step, the logarithm in the 
first term on the right-hand side
of Eq.~(\ref{ferdet}) is expressed as a power series:
\begin{equation}
\mbox{Tr} \ln( G_0^{-1}-\Sigma^{(0)})=\mbox{Tr} \ln( G_0^{-1})-
\sum_{n=1}^{\infty} \frac{1}{n} \mbox{Tr} (G_0 \Sigma^{(0)})^n\,.
\label{logseries}
\end{equation}
The first term represents the free part and 
will enter the path integral as a multiplicative  
normalization constant. Our task is to
take the trace over spin and momentum indices, and obtain a
closed expression for general $n$, so that the infinite series
can be re-summed. 
It is useful to write down explicitly 
the trace over the matrix product in the
above equation:
\begin{eqnarray}
\mbox{Tr} \left( G_0 \Sigma^{(0)} \right)^n  &=& 
 \left(\frac{g}{2}\right)^n 
\sum_{\stackrel{\alpha_1 \ldots \alpha_n} {k_1 \ldots k_n}} 
g_0(k_1) \sigma_{\alpha_1 \alpha_2}^{a_1} 
\Bigl[ n_-^{a_1}(k_1-k_2-k_s)+n_+^{a_1}(k_1-k_2+k_s)
\Bigr]  \nonumber \\
& &\mbox{\hspace*{2cm}} g_0(k_2)\sigma_{\alpha_2 \alpha_3}^{a_2}
\Bigl[
n_-^{a_2}(k_2-k_3-k_s)+n_+^{a_2}(k_2-k_3+k_s )\Bigr] \nonumber \\
& &\mbox{\hspace*{2cm}} \vdots\nonumber \\
& &\mbox{\hspace*{2cm}} g_0(k_n)\sigma_{\alpha_n\alpha_1}^{a_n}
\Bigl[n_-^{a_n}(k_n-k_1-k_s )+n_+^{a_n}(k_n-k_1+k_s)
\Bigr]. 
\label{matrprod}
\end{eqnarray}
Multiplying out the brackets 
leads to a sum 
\begin{eqnarray}
\sum_{r_1=-,+}\! \ldots \! \sum_{r_n=-,+}& & g_0(k_1)
\sigma_{\alpha_1 \alpha_2}^{a_1}n_{r_1}^{a_1}(k_1-k_2+r_1k_s) \ldots
g_0(k_n)\sigma_{\alpha_n\alpha_1}^{a_n}n_{r_n}^{a_n}(k_n-k_1+r_n
k_s).\nonumber \\
& &
\end{eqnarray}
Since we are taking the trace, the sum of all momentum transfers to
the spin fields must be zero, so that
$k_{n+1} \stackrel{!}{=} k_1$. 
Since $k_s=(0,\vec{k}_s)$ (where $\vec{k}_s=\vec{Q}+\delta
\vec{Q}$, see Eqs.~(\ref{ks})--(\ref{dku})) is
incommensurate, this
can be fulfilled only if $n$ is
even and the number of $r_i=(-)$ equals the number of
$r_i=(+)$.
Consequently, the summmation $\sum_{r_1\ldots r_n}$ over all
configurations
of the $\{r_i\}$ contains ${{n} \choose {n/2}}$ terms,
namely all permutations of $(n/2)$ $\vec{n}_-$-fields with $(n/2)$
$\vec{n}_+$-fields.
Since $n$ is even, the trace in spin space over the string of
Pauli matrices can be carried out using the trace reduction 
formula \cite{veltm}:
\begin{equation}
\mbox{Tr}\left(\sigma^{a_1}\sigma^{a_2} \ldots \sigma^{a_n}
\right) = 2\sum_P\!' (-1)^P \delta^{a_1a_{i_2}} \ldots
 \delta^{a_{i_{n-1}}a_{i_n}}\,, \label{tracered}
 \end{equation}
where $P$ is the permutation 
\begin{equation}
P={\!{2\,\,3\,\,4\,\ldots \,n} \choose {i_2\,i_3\,i_4\ldots
i_n}}\,,
\end{equation}
and the sum $\sum_P'$  includes permutations between different
index pairs only. In order to perform the gradient expansion of the
expression $\mbox{Tr} (G_0 \Sigma^{(0)})^n$,
the momenta in the arguments of the fields are redefined \cite{dipapp}
in such a way that the free fermionic
propagators $g_0$  appearing in Eq.~(\ref{matrprod})
can be expanded in powers of the momentum transfer to the spin
field. We obtain
\begin{eqnarray}
\mbox{Tr} \left( G_0 \Sigma^{(0)}\right)^n&=&
2 \left(\frac{g}{2}\right)^n\!\sum_{q_1 \ldots
q_{n-1}}\sum_{r_1\ldots r_n}\sum_P\!'
(-1)^P \delta^{a_1a_{i_2}}\ldots \delta^{a_{i_{n-1}}a_{i_n}}
\nonumber \\
& &\sum_k g_0(k)g_0(k-q_1+r_1k_s)\ldots
g_0(k-q_{n-1}+{\textstyle \sum_{i=1}^{n-1}} r_i\,k_s)\times \nonumber
\\
& &n^{a_1}_{r_1}(q_1) n^{a_2}_{r_2}(q_2-q_1)
\ldots n^{a_n}_{r_n}(-q_{n-1})
\end{eqnarray}
We abbreviate the product of propagators as
\begin{eqnarray}
\label{Pi}
\Pi(q_1\ldots q_{n-1};r_1\ldots r_n) &=& \sum_k
g_0(k)g_0(k-q_1+r_1k_s)\ldots
g_0(k-q_{n-1}+{\textstyle \sum_{i=1}^{n-1}} r_i\,k_s)\,.
\nonumber \\
& &
\end{eqnarray}
Note that the form of the function $\Pi $ depends on the set of
parameters $\{r_i\}$.
The $q$'s are small, so we can expand the function
$\Pi$ around $k$.
It should be stressed that only the
momenta exchanged with the spin fields are small but not their
values for a given fermionic state. By Fourier transformation of
the spin fields back into direct space and subsequent 
integration by parts, one obtains from an expansion up to
$\cal O$($q^2$) first and second order spatial and
temporal derivatives of the spin fields.
Defining
\begin{eqnarray}
\Gamma_{i}^{\mu}\,&=&\int\! {\rm d}^2x {\rm d}\tau\,\, \sum_{l=1}^{i} 
		   \left[ n^{a_1}_{r_1}(x)\ldots
\partial_{\mu} n^{a_l}_{r_l}(x)\ldots n^{a_i}_{r_i}(x) \right]
n^{a_{i+1}}_{r_{i+1}}(x)\ldots  n^{a_n}_{r_n}(x)\,,\label{gammai}\\
\Gamma_{i}^{\mu \nu}&= &\int\! {\rm d}^2x {\rm d}\tau \sum_{l,m=1}^{i}
		     \left[  n^{a_1}_{r_1}(x) \ldots
\partial_{\mu} n^{a_l}_{r_l}(x)\ldots \partial_{\nu}
n^{a_m}_{r_m}(x)\ldots
n^{a_i}_{r_i}(x) \right] n^{a_{i+1}}_{r_{i+1}}(x)\ldots
n^{a_n}_{r_n}(x)\,, \label{gammaii}\nonumber \\
& & \\
\Gamma_{ij}^{\mu \nu}&=&
		      \int\! {\rm d}^2x {\rm d}\tau
		      \, \Bigl\{ \sum_{l,m=1}^{i}\left[ 
		      n^{a_1}_{r_1}(x)\ldots \partial_{\mu} 
		      n^{a_l}_{r_l}(x)\ldots \partial_{\nu}
n^{a_m}_{r_m}(x)\ldots n^{a_i}_{r_i}(x) \right]  n^{a_{i+1}}_{r_{i+1}}
(x)\ldots
n^{a_n}_{r_n}(x)\nonumber \\
& &+\sum_{l=1}^{i} \sum_{m=i+1}^{j}\left[  n^{a_1}_{r_1}(x)\ldots
\partial_{\mu} n^{a_l}_{r_l}(x)\ldots n^{a_i}_{r_i}(x) \right] \bigr[
 n^{a_{i+1}}_{r_{i+1}}(x)\ldots\partial_{\nu} n^{a_m}_{r_m}
 (x)\ldots n^{a_j}_{r_j}(x)
\bigl] \nonumber \\
 & & \hspace*{2.1cm} n^{a_{j+1}}_{r_{j+1}}(x)\ldots
 n^{a_n}_{r_n}(x)\Bigr\} \label{gammaij} \,,
 \end{eqnarray}
we have
\begin{eqnarray}
\mbox{Tr} \left( G_0 \Sigma^{(0)} \right)^n&=&
2 \left(\frac{g}{2}\right)^n \frac{1}{a^2}
\sum_{r_1 \ldots r_n} \sum_P\!' (-1)^P \delta^{a_1a_{i_2}}
\ldots \delta^{a_{i_{n-1}}a_{i_n}}  \nonumber \\ 
& & \Biggl\{\, \sum_{i=1}^{n-1}{\frak i}\,
\frac{\partial \Pi}{\partial
q_i^{\mu}}\Bigr|_{q_i=0}\,\Gamma_{i}^{\mu}
+\frac{1}{2}\sum_{i=1}^{n-1}
(-1)\,\frac{\partial^2\Pi}
{\partial q_i^{\mu}\partial q_i^{\nu}}\Bigr|_{q_i=0}\,
\Gamma_{i}^{\mu \nu}+\nonumber \\
& &+\frac{1}{2}\sum_{\stackrel{i,j=1}{i\neq
j}}^{n-1}(-1) \,
\frac{\partial^2 \Pi}{\partial q_i^{\mu}\partial
q_j^{\nu}}\Bigr|_{q_i=0}\,\Gamma_{ij}^{\mu \nu}\Biggr\}\,.
\label{trn}
\end{eqnarray}
Here $\partial_{\mu}$ is an abbreviation for $\partial /\partial
x^{\mu}$.
Analogously to our notation for the 3-momentum,
in which $k=(\epsilon_n,\vec{k})$, we define
$x=(\tau,\vec{x})$,  so that the indices $\mu$ and $\nu$ in
Eqs.~(\ref{gammai})--(\ref{trn}) run over three values.
In order to avoid confusion with the summation index
$i$, we use ${\frak i}$ to represent $\sqrt{-1}$ 
in this section. The boundary terms
resulting from the integration by parts, which we performed to
arrive at Eq.~(\ref{trn}), vanish because the field
$\vec{n}(x)$ has been required to be constant in infinity,
which is a natural assumption in a low-temperature approach.
As is evident from Eq.~(\ref{trn}), the Kronecker deltas
$\delta^{a_ia_j}$ generate a pairwise contraction
of the fields and their derivatives
into inner products.
Carrying out the sum $\sum_P'$, we sum over all
possible pairwise contractions.
In the derivation of Eq.~(\ref{trn}), the local part containing
the zeroth order term in the expansion of $\Pi$ has been
discarded since it contains no derivatives and  is therefore
constant due to the constraint, Eq.~(\ref{constr}).
Several types of inner products are encountered in
Eq.~(\ref{trn}):
\begin{eqnarray}
n_+^a\,n_+^a&=&n_-^a\,n_-^a\;\;\;\;\;\,=0\,, \label{constr1}\\
(\partial_{\mu} n_+^a)n_+^a&=&(\partial_{\mu} n_-^a)n_-^a=0 \,,
\label{constr4}\\
n_+^a\,n_-^a&=&2\label{constr2}\,,\\
(\partial_{\mu} n_-^a)n_+^a&=&-(\partial_{\mu} n_+^a)n_-^a=2\,
{\frak i}\,
(\partial_{\mu}n_1^a)n_2^a \,,\label{constr3} \\
(\partial_{\mu} n_-^a) \,(\partial_{\nu}n_+^a)&=&
(\partial_{\mu} n_1^a)\,(\partial_{\nu} n_1^a )
+(\partial_{\mu} n_2^a)\,(\partial_{\nu} n_2^a)\,.
\label{constr5}
\end{eqnarray}
Pair contractions of the type
$(\partial_{\nu}n_-^a) \,(\partial_{\nu} n_-^a)$ need not be
considered since they imply a contraction  of the type
$n_+^an_+^a$ (because the number of fields is even),
 which gives zero according  to Eq.~(\ref{constr1}).
 From Eqs.~(\ref{constr1})--(\ref{constr5}), it is clear that
 one gets nonzero contributions only from those contractions
 $\delta^{a_ia_j}$ for which $r_i=-r_j$. Since 
there are $(\frac{n}{2})!$ possibilities  for such contractions
producing inner products of $(\frac{n}{2})$ $\vec{n}_+$-fields  
with $(\frac{n}{2})$ $\vec{n}_-$-fields, the
 sum $\sum_P'$ in $X_n^I$ has $(\frac{n}{2})!$ terms.

We will now outline the combinatorial analysis which we have performed 
in order to permit an explicit evaluation of the sums appearing in
Eqs.~(\ref{gammai}) -- (\ref{trn}) to arbitrary order in $n$. This analysis
departs considerably from previous work in the antiferromagnetic case 
\cite{muze,mulett}. We denote the contribution of the first-order
derivatives to \mbox{Tr}$(G_0 \Sigma^{(0)})^n$ by 
\begin{equation}
\label{Xi}
X_n^I = 2 \left(\frac{g}{2}\right)^n \frac{1}{a^2}
\sum_{r_1\ldots r_n} \sum_P\!'
(-1)^P\delta^{a_1a_{i_2}}
 \ldots
 \delta^{a_{i_{n-1}}a_{i_n}}\sum_{i=1}^{n-1}
  {\frak i} \,\frac{\partial \Pi}{\partial q_i^{\mu
 }}\Bigr|_{q_i=0}\, 
 \Gamma_{i}^{\mu}\,,
 \end{equation}
 where $\Gamma_{i}^{\mu}$ was defined in Eq.~(\ref{gammai}).
We will show  that the only nonvanishing
contributions to the sum $\sum_{r_1 \ldots r_n}$  in $X_n^I$
come from strictly alternating configurations of the set
$\{r_i\}$, i.e.~those configurations which fulfill
$r_{i+1}=-r_i$.

Let us consider a {\em nonalternating} configuration of the
$\{r_i\}$, where at least two $n_+$-terms (and  consequently two
$n_-$-terms) are nearest neighbours, i.e.~we assume $r_i=r_{i+1}$
for some $i$. One of the
possible contractions connecting only 
$n_+$-fields with $n_-$-fields is:
\begin{equation}
\left[ \ldots n^{a_i}_- \ldots n^{a_j}_+ n^{a_{j+1}}_+ \ldots
n^{a_k}_- \ldots \right]\delta^{a_ia_j} \delta^{a_{j+1}a_k}\,.
\label{p1}
\end{equation}
Another contraction connecting only $n_+$-fields with
$n_-$-fields by is obtained by  one permutation:
\begin{equation}
\left[ \ldots n^{a_i}_- \ldots n^{a_j}_+ n^{a_{j+1}}_+ \ldots
n^{a_k}_- \ldots \right]\delta^{a_{i}a_{j+1}} \delta^{a_ja_k}\,.
\label{p2}
\end{equation}
Two sets of contractions differing by one permutation carry the
sign  $(-1)^P$, $(-1)^{P+1}$, respectively. Since the 
two terms (\ref{p1}) and (\ref{p2}) are identical except 
for their relative
sign, they add up to zero in the sum $\sum_P'$. 
Since $(\frac{n}{2})!$ is an even number, 
one can find for every nonvanishing
set of contractions another one differing only by a relative
sign given a nonalternating configuration of the
$\{r_i\}$. Thus all terms in the sum $\sum_P'$ add to zero
pairwise for a nonalternating configuration of the $\{r_i\}$.
 From the sum $\sum_{r_1\ldots r_n}$, 
we are left only with the two alternating configurations.

For a strictly alternating configuration of the  $\{r_i\}$, where
$r_{i+1}=-r_i \,\forall\,\, i$, the total number of nonvanishing
terms from the sum $\sum_P'$ is obtained by the following
consideration:
If a pair-contraction $\delta^{a_ia_j}$ with $r_i=-r_j$ 
leaves a nonalternating sequence of the  $\{r_i\}$, all
terms will again add to zero pairwise. 
The only contractions that always leave an
alternating sequence of the  $\{r_i\}$ are contractions between
nearest neighbours or between the first and last member of the
remaining sequence. The number of these contractions is
$2^{(\frac{n}{2}-1)}$, and they carry the same sign.
Thus in Eq.~(\ref{Xi}), the sum $\sum_P' (-1)^P
\delta^{a_1a_{i_2}}\ldots  \delta^{a_{i_{n-1}}a_{i_n}}$ can be
replaced by $2^{(\frac{n}{2}-1)}$ times the contribution of the
zeroth permutation, $\delta^{a_1a_2}\delta^{a_3a_4}\ldots
\delta^{a_{n-1}a_{n}}$. 

After the sums
$\sum_{r_1\ldots r_n}$ and $\sum_P'$ have been discussed, we
still need to carry out the sums
$\sum_{i=1}^{n-1}$ and $\sum_{l=1}^{i}$ 
 (see Eqs.~(\ref{gammai}),(\ref{Xi})) in order to obtain a closed
expression for $X_n^I$.
First we consider the alternating
configuration starting with $r_1=(-)$. 
For this configuration of the $\{r_i\}$ and the
set of contractions $\delta^{a_1a_2}\delta^{a_3a_4}
\ldots \delta^{a_{n-1}a_{n}}$,
every term with $l$ odd gives $(\partial n_-^a)n_+^a$, while
every term with $l$ even gives $(\partial n_+^a)n_-^a$. 
There is always one contraction containing the derivative,
for which we use Eq.~(\ref{constr3}), and $(\frac{n}{2}-1)$
contractions without derivative, for which we use
Eq.~(\ref{constr2}). This leads to
\begin{eqnarray}
\sum_{l=1}^{i} \left[ n^{a_1}_{r_1}\ldots \partial_{\mu}
n^{a_l}_{r_l}\ldots n^{a_i}_{r_i} \right]
n^{a_{i+1}}_{r_{i+1}}\ldots  n^{a_n}_{r_n}&=&
\sum_{l=1}^{i} (-1)^{(l-1)}\, 2\,{\frak i}\,
(\partial_{\mu} n_1^a)n_2^a\;
2^{(\frac{n}{2}-1)}=\nonumber \\
&=&2^{\frac{n}{2}}{\frak i}\,(\partial n_1^a)n_2^a \times
\left\{\begin{array}{l}
\mbox{1 for $i$ odd,}\\
\mbox{0 for $i$ even.}
\end{array}
\right.
\end{eqnarray}
For $i$ odd, the first derivatives of the product of propagators
$\Pi$ (defined in Eq.~(\ref{Pi})) for an alternating
configuration starting with $r_1=(-)$ acquire the form
\begin{eqnarray}
\frac{\partial \Pi^{-+}}{\partial q^{\mu}_{i_{odd}}}\Bigr|_{q_i=0}&=&
\sum_k\; \bigl[ g_0(k) \bigr]^{\frac{n}{2}} \bigl[
g_{0,-}(k)\bigr]^{\frac{n}{2}-1} \frac{\partial
g_{0,-}(k-q)}{\partial q^{\mu}}\Bigr|_{q=0}= \nonumber\\
 &=&\frac{2}{n}\sum_k\; \bigl[ g_0(k)
\bigr]^{\frac{n}{2}}\frac{\partial }{\partial q^{\mu}}\bigl[
g_{0,-}(k-q)\bigr]^{\frac{n}{2}}\Bigr|_{q=0}\,.
\label{Pmp}
\end{eqnarray}
We define
\begin{equation}
g_{0,\pm}(k) =g_0(k\pm k_s)\,.
\end{equation}
Since $\partial \Pi^{-+}/\partial q^{\mu}_{i_{odd}}|_{q_i=0}$
has the same form for all $i$ odd, the summation over $i$ can 
be carried out:
$\sum_{i=1,i\,odd}^{n-1} 1=\frac{n}{2}$.
Inserting these results into Eq.~(\ref{Xi}),
and recalling that the sum $\sum_P'$ over the permutations is
replaced by $2^{(\frac{n}{2}-1)}$ times the contribution of the
zeroth permutation,  we obtain 
\begin{eqnarray}
X^{I}_{n(-+)}&=& 2 \,\left(\frac{g}{2}\right)^n
	  2^{(\frac{n}{2}-1)}\frac{{\frak i}}{a^2}\,\frac{n}{2}\,
	  \frac{\partial \Pi^{-+}}{\partial
q^{\mu}_{i_{odd}}}\Bigr|_{q_i=0} \int {\rm d}^2x {\rm d}\tau\, 2^{\frac{n}{2}} \,
{\frak i}\, 
(\partial_{\mu} n_1^a)n_2^a=\nonumber \\
&=&-\frac{n}{2a^2}\,g^n\,\frac{\partial \Pi^{-+}}{\partial
q^{\mu}_{i_{odd}}}\Bigr|_{q_i=0} \int {\rm d}^2x {\rm d}\tau\, (\partial_{\mu}
n_1^a)n_2^a\,.
\label{Xmp}
\end{eqnarray}
For our fermion determinant, Eq.~(\ref{logseries}), we need
$-\sum_{n=1}^{\infty} \frac{1}{n} X_{n(-+)}^{I}$.
We insert Eq.~(\ref{Pmp}) into Eq.~(\ref{Xmp}) and 
re-sum the power series in $n$:
\begin{eqnarray}
-\! \sum_{\stackrel{n=2}{n\,even}}^{\infty} \frac{1}{n} 
X^{I}_{n(-+)}
&=&
\sum_{n=1}^{\infty}\,\frac{g^{2n}}{2n}\,\frac{1}{a^2}\sum_k \,\bigl[
g_0(k)\bigr]^n \frac{\partial }{\partial q^{\mu}}
\bigl[g_{0,-}(k-q)\bigr]^n \Bigr|_{q=0}\int {\rm d}^2x {\rm d}\tau\,
(\partial_{\mu} n_1^a)n_2^a=  \nonumber \\
&=& -\frac{1}{2a^2} \,\frac{\partial }{\partial q^{\mu}} 
\sum_k \ln \Bigl[
1-g^2g_0(k)g_{0,-}(k-q)\Bigr]\Bigr|_{q=0} \int {\rm d}^2x {\rm d}\tau\, 
(\partial_{\mu}
n_1^a)n_2^a\,.  \nonumber \\
& &
\end{eqnarray}
Adding also the contribution $X_{n(+-)}^{I}$ from the 
alternating sequence starting with $r_1=(+)$ and introducing 
\begin{eqnarray}
\Phi^{\pm}_1(q)&= &
\sum_k \ln( 1-g^2g_0(k)g_{0,\pm}(k-q))\,,
\label{Phi}
\end{eqnarray}
we obtain
\begin{equation}
-\sum_{n=2}^{\infty} \frac{1}{n}
X_n^I= \frac{1}{2a^2} 
\frac{\partial}{\partial q^{\mu}} \Bigl(\Phi^+_1(q)-\Phi^-_1(q)
\Bigl)\Bigr|_{q=0}
\int {\rm d}^2x {\rm d}\tau \,(\partial_{\mu}n_1^a)n_2^a \,.
\end{equation}
The sum $\sum_k$ in our expressions for $\Phi^{\pm}_1(q)$
denotes a sum $\sum_{\epsilon_n,\vec{\scriptstyle k}}$.
By shifting the summation index $\vec{k} \rightarrow -\vec{k}$ 
in $\Phi^+_1(q)$ and using the symmetry of the free propagator, 
Eq.~(\ref{g0sym1}), it can be shown that
\begin{eqnarray}
\frac{\partial \Phi^+_1(q)}{\partial q^0}\Bigr|_{q=0}&=&
\frac{\partial \Phi^-_1(q)}{\partial q^0}\Bigr|_{q=0}\,, \nonumber \\
\frac{\partial \Phi^+_1(q)}{\partial q^{1,2}}\Bigr|_{q=0}&=&
-\frac{\partial \Phi^-_1(q)}{\partial q^{1,2}}\Bigr|_{q=0}\,,
\label{symm}
\end{eqnarray}
such that the time component of the derivative drops out and 
only the space components remain:
\begin{equation}
-\sum_{n=2}^{\infty} \frac{1}{n} X_n^I=
\sum_{\mu=1}^2 \frac{1}{a^2}
\frac{\partial \Phi^-_1(q)}{\partial q^{\mu}}
\Bigr|_{q=0}\int {\rm d}^2x {\rm d}\tau \,\vec{n}_1\cdot
\partial_{\mu}\vec{n}_2 \,. \label{1stres}
\end{equation}
For $\delta \vec{Q} = 0$, i.e.~a pure antiferromagnetic
configuration, the first-order contributions vanish. 
We will discuss the relevance of this term for the resulting 
field theory at a later stage, after we have also obtained
a contribution linear in the derivatives when we take the
continuum limit of the Heisenberg part of the action.

We now turn to the second-order terms in the expression (\ref{trn}).
The contribution of the quadratic derivatives is denoted by 
\begin{equation}
\label{Xii}
X_n^{II} =  \left(\frac{g}{2}\right)^n \frac{1}{a^2}
\sum_{r_1\ldots r_n}
\sum_P\!'\, (-1)^P\delta^{a_1a_{i_2}}  \ldots
\delta^{a_{i_{n-1}}a_{i_n}} \sum_{i=1}^{n-1}
(-1)\frac{\partial^2\Pi}{\partial q_i^{\mu}\partial q_i^{\nu}}
\Bigr|_{q_i=0}\, \Gamma_{i}^{\mu \nu} 
\end{equation}
(see Eq.~(\ref{gammaii}) for $\Gamma_i^{\mu \nu}$).
We proceed similarly as in the case of the first-order
derivatives, and investigate first whether for some special
configurations of the $\{r_i\}$ the terms in the sum over 
all contractions $\sum_P'$ add to zero. 
It will turn out that in the case of the quadratic
derivatives
not only the strictly alternating configurations of the 
$\{r_i\}$ give a nonvanishing contribution, but also 
configurations where the alternation is broken once.
Again, in a nonalternating sequence one can find two possible
sets of pair-contractions, differing by one permutation,
that connect only $(+-)$-pairs and
carry the signs $(-1)^P$ and $(-1)^{P+1}$, respectively.  
The corresponding terms will only add to zero if they are identical
except for their signs. 
However, this is not always the case if
we deal with the second-order-derivatives. It may occur that in
one of the two sets the two derivatives are contracted, 
\begin{equation}
\left[ \ldots \partial_{\mu} n^{a_i}_- \ldots \partial_{\nu} n^{a_j}_+ 
\ldots n^{a_{k}}_+ \ldots
n^{a_l}_- \ldots \right]\delta^{a_ia_j} \delta^{a_{k}a_l}\,.
\label{pp1}
\end{equation}
while
in the set  differing by one permutation 
the derivatives are contracted with other
fields, i.e.
\begin{equation}
\left[ \ldots \partial_{\mu} n^{a_i}_- \ldots \partial_{\nu} n^{a_j}_+
\ldots n^{a_{k}}_+ \ldots
n^{a_l}_- \ldots \right]\delta^{a_ia_{k}} \delta^{a_{j}a_l}\,.
\label{pp2}
\end{equation}
While a contraction  of type (\ref{pp1}) gives a term of the
form $[(\partial_{\mu} \vec{n}_1)\cdot(\partial_{\nu} \vec{n}_1 ) 
+(\partial_{\mu} \vec{n}_2)\cdot(\partial_{\nu} \vec{n}_2)]$ 
(see Eq.~(\ref{constr5})), the contraction (\ref{pp2})  yields 
$4(\vec{n}_1\cdot \partial_{\mu}\vec{n}_2 )(\vec{n}_1\cdot
\partial_{\nu}\vec{n}_2 )$ (see
Eq.~(\ref{constr3})). These terms are not identical but 
related through
\begin{equation}
2(\vec{n}_1\!\cdot\! \partial_{\mu}\vec{n}_2 )(\vec{n}_1\!\cdot
\!\partial_{\nu}\vec{n}_2 )
= (\partial_{\mu} \vec{n}_1)\cdot(\partial_{\nu} \vec{n}_1 )
 +(\partial_{\mu} \vec{n}_2)\cdot(\partial_{\nu} \vec{n}_2)-
 (\partial_{\mu} \vec{n}_3)\cdot(\partial_{\nu} \vec{n}_3)
\,. \label{n3rel}
\end{equation}
Whether two terms carrying the signs $(-1)^P$  and $(-1)^{P+1}$
are identical depends on the
position of the derivatives,
which is denoted by $l$ and $m$ (see Eq.~(\ref{gammaii}))
and varied in the sum $\sum_{l,m=1}^i$. 
Thus we will now evaluate
the sum  $\sum_P'$    for all possible positions $l$ and $m$ of the
derivatives and then compute the sum $\sum_{l,m=1}^i$.
Again, we first discuss the alternating configuration starting
with $r_1=(-)$.
The contribution of terms with $l=m$ is 
\begin{eqnarray}
T^{\mu \nu}_{(l=m)}&=&\sum_P\!'\, (-1)^P\delta^{a_1a_{i_2}}\ldots
\delta^{a_{i_{n-1}}a_{i_n}} \int\! {\rm d}^2x {\rm d}\tau\,
\left[  n^{a_1}_{r_1}
\ldots (\partial_{\mu} \partial_{\nu} n^{a_l}_{r_l})\ldots
n^{a_n}_{r_n}\right] \nonumber \\
&=&-2^{(n-2)} 
\sum_{b=1}^2 (\partial_{\mu} \vec{n}_b)\cdot (\partial_{\nu} 
\vec{n}_b) \,. \label{t1}
\end{eqnarray}
For terms with $l \neq m$, where $l$, $m$ are both even or both
odd, we obtain ($z\, \epsilon\,{\Bbb N} \backslash \{0\}$)
\begin{eqnarray}
T^{\mu \nu}_{(l=m\pm 2z)}&=&\sum_P\!'\,
(-1)^P\delta^{a_1a_{i_2}}\ldots
\delta^{a_{i_{n-1}}a_{i_n}} 
\int\! {\rm d}^2x {\rm d}\tau\,\left[  n^{a_1}_{r_1} \ldots 
 (\partial_{\mu}  n^{a_l}_{r_l})\ldots  (\partial_{\nu}
n^{a_{l\pm 2z}}_{r_{l\pm 2z}})\ldots n^{a_n}_{r_n}\right] \nonumber
\\
&=& -2^{(n-2)} \; 2\;(\vec{n}_1\cdot \partial_{\mu}\vec{n}_2
)(\vec{n}_1\cdot
\partial_{\nu}\vec{n}_2 )\,.\label{t2}
\end{eqnarray}
Terms with $l=m \pm 1$ yield
\begin{eqnarray}
T_{(l=m\pm 1)}^{\mu \nu}&=&\sum_P\!' \,
(-1)^P\delta^{a_1a_{i_2}}\ldots
\delta^{a_{i_{n-1}}a_{i_n}}\int\! {\rm d}^2x {\rm d}\tau
\,\bigl[  n^{a_1}_{r_1}
\ldots(\partial_{\mu}
n^{a_l}_{r_l})(\partial_{\nu}n^{a_{l\pm 1}}_{r_{l\pm 1}})\ldots
n^{a_n}_{r_n}\bigr] \nonumber \\
&=&2^{(n-3)}\,\sum_{b=1}^2 (\partial_{\mu}
\vec{n}_b) \cdot (\partial_{\nu} \vec{n}_b)
+2^{(n-3)}\,2\,(\vec{n}_1\cdot \partial_{\mu}\vec{n}_2
)(\vec{n}_1\cdot
\partial_{\nu}\vec{n}_2 )\,. \label{t3}
\end{eqnarray}
Terms with $l \neq m \pm 1$, where $l$ even, $m$ odd
or vice versa, contribute
\begin{eqnarray}
T^{\mu \nu}_{(l=m\pm 2z+1)}
&=&\sum_P\!'\, (-1)^P\delta^{a_1a_{i_2}}\ldots 
\delta^{a_{i_{n-1}}a_{i_n}}
\int\! {\rm d}^2x {\rm d}\tau\,\left[  n^{a_1}_{r_1} \ldots 
 (\partial_{\mu}  n^{a_l}_{r_l})\ldots \right. \nonumber \\
& &\left. \mbox{\hspace*{5cm}}\ldots (\partial_{\nu}
n^{a_{l\pm 2z+1}}_{r_{l\pm 2z+1}})\ldots n^{a_n}_{r_n}\right]\nonumber
\\
&=& 2^{(n-2)}
\; 2\;(\vec{n}_1\cdot \partial_{\mu}\vec{n}_2 )(\vec{n}_1\cdot
\partial_{\nu}\vec{n}_2 )\,. \label{t4}
\end{eqnarray}
We now determine the number $N$ of each of these four types of $T^{\mu
\nu}$-terms in the sum $\sum_{l,m=1}^i$, and denote the result of the
summations $[\sum_{l,m=1}^i  \sum_P']$ by ${\cal T}_i$, 
i.e.~${\cal T}_i=N_{(l=m)}\,T^{\mu \nu}_{(l=m)} 
+N_{(l=m\pm 2z)}\, T^{\mu \nu}_{(l=m\pm 2z)}
+ N_{(l=m \pm 1)}\, T^{\mu \nu}_{(l=m\pm 1)}
+ N_{(l=m \pm 2z +1)} \,T^{\mu \nu}_{(l=m\pm 2z+1)}$.
Using Eq.~(\ref{n3rel}), we obtain that ${\cal T}_i$ does not 
depend on explicitly on $i$, namely,
\begin{eqnarray}
{\cal T}_i(i\!=\! \mbox{odd})&=&-2^{(n-2)} \sum_{b=1}^2
(\partial_{\mu}
\vec{n}_b) \cdot (\partial_{\nu} \vec{n}_b)\,, \nonumber \\
{\cal T}_i(i\!=\!\mbox{even})&=&-2^{(n-2)}(\partial_{\mu}
\vec{n}_3) \cdot (\partial_{\nu} \vec{n}_3)\,. \label{Tevenodd}
\end{eqnarray}
For the alternating configuration starting with $r_1=(-)$ which
we have discussed so far, the second derivative of the product
of propagators from Eq.~(\ref{Pi}) becomes
\begin{eqnarray}
\frac{\partial^2 \Pi^{ -+}}{\partial q_{i_{odd}}^{\mu} \partial
q_{i_{odd}}^{\nu}} \Bigl|_{q_i=0} &=& \sum_k \, g_0(k)^{\frac{n}{2}}
\, g_{0,-}(k)^{\frac{n}{2}-1} 
\frac{\partial^2 g_{0,-}(k-q)}{\partial q^{\mu} \partial q^{\nu}} 
\Bigl|_{q=0} \,, \nonumber
\\
\frac{\partial^2 \Pi^{ -+}}{\partial q_{i_{even}}^{\mu} \partial
q_{i_{even}}^{\nu}} \Bigl|_{q_i=0} &=& \sum_k \,
g_0(k)^{\frac{n}{2}-1}\,
g_{0,-}(k)^{\frac{n}{2}}\,
\frac{\partial^2 g_{0}(k-q)}{\partial q^{\mu}
\partial q^{\nu}} \Bigl|_{q=0}  \,,
\end{eqnarray}
so that the sum $\sum_{i=1}^{n-1}$ can be carried out.
Adding the contribution of the alternating configuration starting
with $r_1=(+)$, we obtain for the contribution of
the two alternating configurations to the sum $\sum_{r_1 \ldots
r_n}$ in Eq.~(\ref{Xii}):
\begin{eqnarray}
X^{II}_{n(alt)} &=& \left(\frac{g}{2}\right)^n
\,\frac{2^{(n-2)}}{a^2}\, \Bigl[\, \sum_k\, g_0(k)^{\frac{n}{2}}\,
g_{0,-}(k)^{\frac{n}{2}-1} 
\frac{\partial^2 g_{0,-}(k-q)}{\partial q^{\mu}
\partial q^{\nu}} \Bigl|_{q=0} \nonumber \\
& &+\sum_k \,
g_0(k)^{\frac{n}{2}-1}\,g_{0,-}(k)^{\frac{n}{2}}\,\frac{\partial^2
g_{0}(k-q)}{\partial q^{\mu}\partial q^{\nu}} \Bigl|_{q=0}\,
\Bigl]\times \nonumber \\
& & \int \!{\rm d}^2x {\rm d}\tau 
\Bigl[\, \frac{n}{2} \sum_{b=1}^2 (\partial_{\mu}
\vec{n}_b) \cdot (\partial_{\nu} \vec{n}_b)+ (\frac{n}{2}-1)\,
(\partial_{\mu} \vec{n}_3) \cdot (\partial_{\nu} \vec{n}_3)
\Bigr]\,.\label{XnIIalt} 
\end{eqnarray}
For a nonalternating
configuration, the terms  $T^{\mu \nu}_{(l=m)}$ and 
$T^{\mu \nu}_{(l=m \pm 2z)}$ vanish. The terms 
$T^{\mu \nu}_{(l=m\pm 1)}$ and $T^{\mu \nu}_{(l=m\pm 2z+1)}$ can be
nonvanishing when the contraction of the two derivatives is
such that the remaining sequence is alternating, i.e.~when the
contraction of the derivatives `repairs' the defects in the alternation.
However, such terms can still 
add to zero in the sum $\sum_{l,m=1}^i$. A term with fixed $l$, $m$,
where the contraction of the two derivatives restores the
perfect alternation of the remaining sequence, 
adds to zero e.g.~with the
term $l$, $m+1$, where the `repairing' contraction differs by one
permutation and thus carries the opposite sign. Therefore, from both 
summations  $\sum_{l,m=1}^i$ and $\sum_P'$, we will get a nonzero
contribution only if just one single term, 
where the contraction of the derivatives restores the 
alternation of the remaining sequence, occurs in
$\sum_{l,m=1}^i$. This happens precisely if the 
alternation of the configuration is broken only once. 
These configurations will be denoted by `one-kink configurations' in
the following. 

We proceed with the discussion of the summations $\sum_{i=1}^{n-1}$,
 $\sum_{l,m=1}^{i}$, $\sum_P'$. 
We denote the site after which the kink occurs by ${\frak k}$.
 From the sum  $\sum_{i=1}^{n-1}$ only
 the term with $i\!=\!{\frak k}$ contributes, and 
from the sum $\sum_{l,m=1}^{{\frak k}}$ only the 
 term with $l\!=\!1$, $m\!=\!{\frak k}$ (or vice versa) is
 nonvanishing. 
Further,  for a one-kink configuration ${\frak k}$ is always
even. The result of the 
summations $[\sum_{l,m=1}^{i}\sum_P']$ is again independent of $i$:
\begin{equation}
{\cal T}_i^{kink} = 2^{(n-2)}(\partial_{\mu}
\vec{n}_3) \cdot (\partial_{\nu} \vec{n}_3)\,. \label{Tikink}
\end{equation}
In order to carry out  the sum over all 
possible one-kink configurations,
 we need the form of the function $\partial^2 \Pi^{kink}/
\partial q^{\mu}_{{\frak k}}  
\partial q^{\nu}_{{\frak k}} |_{q=0}$ for a 
general one-kink configuration of length $n$.
If it starts with $r_1=(-)$, we have
\begin{equation}
\frac{\partial^2 \Pi^{kink}_{-}}{\partial q^{\mu}_{{\frak k}}  \partial
q^{\nu}_{{\frak k}}} \Bigl|_{q_i=0}=\sum_k g_{0,-}(k)^{\frac{{\frak
k}}{2}} \,
 g_{0,+}(k)^{\frac{n}{2}-\frac{{\frak k}}{2}} \,
 g_0(k)^{\frac{n}{2}-1}\,
 \frac{\partial^2 g_0(k-q)}{\partial q^{\mu} \partial q^{\nu}} 
 \Bigl|_{q=0}\,.
\end{equation}
We add the configurations starting with $r_1=(-)$ and $r_1=(+)$ 
and perform the sum $\sum_{{\frak k}=2, {\frak k}\, even}^{n-2}$ over all
possible one-kink configurations. We 
find for the contribution of the one-kink
configurations to the sum $\sum_{r_1 \ldots r_n}$ in
Eq.~(\ref{Xii}): 
\begin{eqnarray}
X^{II}_{n(kink)}&=&\left(\frac{g}{2}\right)^n \,\frac{2^{(n-1)}}{a^2}
\;\Biggl\{\,\sum_k\,g_0(k)^{\frac{n}{2}-1}\,
\frac{\Bigl[g_{0,+}(k)^{\frac{n}{2}-1}
-g_{0,-}(k)^{\frac{n}{2}-1}\Bigr]}{\Bigl[
g_{0,+}(k)^{-1}-g_{0,-}(k)^{-1}\Bigr]} \nonumber \\
& & \frac{\partial^2  g_0(k-q)}{\partial q^{\mu} \partial q^{\nu}} 
\Bigl|_{q=0}\, \Biggl\}\;
\int \!{\rm d}^2x {\rm d}\tau \,(\partial_{\mu}
\vec{n}_3)\! \cdot\! (\partial_{\nu}
\vec{n}_3)\,.  \label{XnIIkink}
\end{eqnarray}
The contribution of the mixed derivatives to Tr$(G_0
\Sigma^{(0)})^n$ from Eq.~(\ref{trn}) is 
\begin{eqnarray}
X_n^{IJ}=  \left(\frac{g}{2}\right)^n \frac{1}{a^2} \sum_{r_1\ldots
r_n}\sum_P\!' (-1)^P\delta^{a_1a_{i_2}}
\ldots\delta^{a_{i_{n-1}}a_{i_n}} 
\sum_{\stackrel{i,j=1}{i\neq j}}^{n-1}(-1) 
\frac{\partial^2 \Pi}{\partial q_i^{\mu}\partial q_j^{\nu}}
\Bigr|_{q_{i,j}=0}\,\Gamma_{ij}^{\mu \nu}\,, \label{Xij}
\end{eqnarray}
where $\Gamma_{ij}^{\mu \nu}$ is given by Eq.~(\ref{gammaij}).
As we will show,  nonvanishing contributions to the mixed
derivatives originate from the strictly alternating
configurations of the $\{r_i\}$  as well as from 
configurations where the alternation is broken up to two
times. We will again study the case of the alternating
configurations first.
As can be seen from Eq.~(\ref{gammaij}), $\Gamma_{ij}^{\mu \nu}$
falls into two parts, the sum $\sum_{l,m=1}^{i}$ and the sum
$\sum_{l=1}^{i} \sum_{m=i+1}^{j}$. The result of the  
summations $[\sum_{l,m=1}^{i} \sum_P']$ is
independent of the value that $j$ takes
and its contribution is given by ${\cal T}_{i}$ in 
Eq.~(\ref{Tevenodd}). The terms occurring in the sum
$\sum_{l=1}^{i} \sum_{m=i+1}^{j}$ are
$T^{\mu \nu}_{(l=m\pm 2z)}$, $T_{(l=m\pm 1)}^{\mu \nu}$ and
$T^{\mu \nu}_{(l=m\pm 2z+1)}$, whose
contributions are given in Eqs.~(\ref{t2}), (\ref{t3}),
and (\ref{t4}). 

We determine the number $M$ of each of these three types of terms
in the sum $\sum_{l=1}^{i} \sum_{m=i+1}^{j}$ for every
value of $i$ and $j$, and denote the result of the summations
$[(\sum_{l,m=1}^{i}+\sum_{l=1}^{i} \sum_{m=i+1}^{j}) \sum_P']$ by
${\cal T}_{ij}$, i.e.~ ${\cal T}_{ij}= {\cal T}_{i}+ M_{(l=m\pm 2z)}\,
T^{\mu \nu}_{(l=m\pm 2z)} + M_{(l=m \pm 1)}\, T^{\mu
\nu}_{(l=m\pm 1)} 
+ M_{(l=m \pm 2z +1)} \,T^{\mu \nu}_{(l=m\pm 2z+1)}$.
We find 
\begin{eqnarray}
{\cal T}_{ij}( i=\mbox{odd},\, j=\mbox{even})&=&
-2^{(n-3)}(\partial_{\mu}
\vec{n}_3) \cdot (\partial_{\nu} \vec{n}_3)\,, 
\nonumber \\
{\cal T}_{ij}( i=\mbox{even},\, j=\mbox{odd})&=&
-2^{(n-3)}(\partial_{\mu} \vec{n}_3) \cdot (\partial_{\nu}
\vec{n}_3)\,,
\nonumber \\
{\cal T}_{ij}( i=\mbox{even},\, j=\mbox{even})&=&
-2^{(n-3)}(\partial_{\mu}
\vec{n}_3) \cdot (\partial_{\nu} \vec{n}_3)\,,
\nonumber \\
{\cal T}_{ij}( i=\mbox{odd},\, j=\mbox{odd})&=&
2^{(n-3)}(\partial_{\mu}\vec{n}_3) \cdot (\partial_{\nu}
\vec{n}_3) - 2^{(n-2)} {\displaystyle \sum_{b=1}^2}
(\partial_{\mu}
\vec{n}_b) \cdot (\partial_{\nu} \vec{n}_b)\,. 
\end{eqnarray}
In order to perform the summation $\sum_{i,j=1,\,i\neq
j}^{n-1}$, we determine the second derivatives of the product of
propagators from Eq.~(\ref{Pi}) for an alternating
configuration starting with $r_1=(-)$,
\begin{eqnarray}
\frac{\partial^2 \Pi^{ -+}}{\partial q_{i_{odd}}^{\mu}
\partial q_{j_{even}}^{\nu}} \Bigl|_{q_{i,j}=0} &=&
\frac{\partial^2 \Pi^{ -+}}{\partial q_{i_{even}}^{\mu} \partial
q_{j_{odd}}^{\nu}} \Bigl|_{q_{i,j}=0} =
\sum_k \,
g_0(k)^{\frac{n}{2}-1}   \, g_{0,-}(k)^{\frac{n}{2}-1}\nonumber
\\
& &\mbox{\hspace*{4.2cm}}\frac{\partial g_0(k-q)}{\partial 
q^{\mu}}\Bigl|_{q=0}
\,\frac{\partial g_{0,-}(k-q)}{\partial q^{\nu}} \Bigl|_{q=0}\,,
\nonumber \\
\frac{\partial^2 \Pi^{ -+}}{\partial q_{i_{even}}^{\mu} \partial
q_{j_{even}}^{\nu}} \Bigl|_{q_{i,j}=0} &=& \sum_k \,
g_0(k)^{\frac{n}{2}-2}   \, g_{0,-}(k)^{\frac{n}{2}} \,
\frac{\partial g_0(k-q)}{\partial q^{\mu}}
\Bigl|_{q=0}\,\frac{\partial g_{0}(k-q)}{\partial q^{\nu}}
\Bigl|_{q=0}\,,
\nonumber \\
\frac{\partial^2 \Pi^{ -+}}{\partial q_{i_{odd}}^{\mu}\partial
q_{j_{odd}}^{\nu}} \Bigl|_{q_{i,j}=0} &=& \sum_k
\,g_0(k)^{\frac{n}{2}}   \, g_{0,-}(k)^{\frac{n}{2}-2} \,
\frac{\partial g_{0,-}(k-q)}{\partial q^{\mu}}
\Bigl|_{q=0}\,\frac{\partial  g_{0,-}(k-q)}{\partial q^{\nu}}
\Bigl|_{q=0}\,.
\end{eqnarray}
We add the alternating configuration starting with $r_1=(+)$,
and obtain from the summmation $\sum_{i,j=1,\,i\neq j}^{n-1}$
the contribution of the alternating configurations to the sum
$\sum_{r_1 \ldots r_n}$ in Eq.~(\ref{Xij})
\begin{eqnarray}
 X^{IJ}_{n(alt)}&\!=\!&\left( \frac{g}{2}\right)^n
\,\frac{2^{(n-2)}}{a^2}\,\Biggl\{\Bigl[\,\sum_k 
\,g_0(k)^{\frac{n}{2}-2}\, g_{0,-}(k)^{\frac{n}{2}}
\,\frac{\partial g_0(k-q)}{\partial q^{\mu}}\Bigl|_{q=0}
\frac{\partial g_0(k-q)}{\partial q^{\nu}} \Bigl|_{q=0}
\nonumber \\
& &{}+ \sum_k g_0(k)^{\frac{n}{2}}
\,g_{0,-}(k)^{\frac{n}{2}-2}\,\frac{\partial g_{0,-}(k-q)}{\partial
q^{\mu}} \Bigl|_{q=0}\frac{\partial g_{0,-}(k-q)}{\partial
q^{\nu}} \Bigl|_{q=0} \Bigr] \times \nonumber \\
& &\int \! {\rm d}^2x  {\rm d}\tau
\Bigl[  \frac{n}{2}\,\bigl(\frac{n}{2} -1\bigr)
\sum_{b=1}^2 (\partial_{\mu}
\vec{n}_b) \cdot (\partial_{\nu} \vec{n}_b) -
\bigl( \frac{n}{2} -1\bigr)\,(\partial_{\mu} \vec{n}_3) \cdot
(\partial_{\nu} \vec{n}_3)\Bigr] \nonumber \\
& &{} +\Bigl[ \sum_k g_0(k)^{\frac{n}{2}-1}\,
g_{0,-}(k)^{\frac{n}{2}-1}\,\frac{\partial g_{0}(k-q)}{\partial
q^{\mu}} \Bigl|_{q=0}\frac{\partial g_{0,-}(k-q)}{\partial
q^{\nu}} \Bigl|_{q=0} \Bigr] \times \nonumber \\
& &\int \! {\rm d}^2x  {\rm d}\tau \,  
\Bigl[ n \,\bigl( \frac{n}{2} -1\bigr)\,(\partial_{\mu}
\vec{n}_3) \cdot (\partial_{\nu} \vec{n}_3)\Bigr] \Biggr\}\,.
\label{XnIJalt}
\end{eqnarray}
Next we discuss the contribution of the
nonalternating configurations to the mixed derivatives. 
For nonalternating
configurations the terms $T^{\mu \nu}_{(l=m)}$ and 
$T^{\mu \nu}_{(l=m\pm 2z)}$ vanish. 
 As in the case of the quadratic derivatives, the terms
$T_{(l=m\pm 1)}^{\mu \nu}$ and $T^{\mu \nu}_{(l=m\pm 2z+1)}$
are nonzero
if the contraction of $l$ and $m$ restores the
alternation of the remaining sequence. For the mixed derivatives
considered now, this is possible if the alternation of the
configuration is broken not more than two times. We will first
evaluate the case of the one-kink configurations.

The first sum in $\Gamma_{ij}^{\mu \nu}$ (see
Eq.~({\ref{gammaij})), $\sum_{l,m=1}^{i}$,
has been discussed for one-kink
configurations already, so that we just recall the result for
${\cal T}_{i}^{kink}$, Eq.~({\ref{Tikink}). In order to
evaluate the sum $\sum_{l=1}^{i} \sum_{m=i+1}^{j}$, we first
take $i<j$. The number of the site after which the kink occurs
is again denoted by ${\frak k}$. A contraction of the
derivatives can restore the alternation of the remaining
sequence only if $l\!=\!1 \wedge m\!=\!{\frak k}$ or $l\!=\!1 
\wedge m\!=\!( {\frak k}+1)$. 
The sum $\sum_{i,j=1,i \neq j}^{n-1}$ only yields 
nonzero  contributions for every $i$ with $i<j$ if
$j={\frak k}$ (where only the term with $l\!=\!1 \wedge m\! =\!
m_{max}\! =\! {\frak k}$ contributes to $\sum_{l=1}^{i}
\sum_{m=i+1}^{j} \sum_P'$),
and for every $j$ with $j>i$ if  $i={\frak k}$ (where only the
terms with $l\!=\!1 \wedge m\! =\! m_{min}\! = \! ({\frak k}+1)$
contributes to $\sum_{l=1}^{i} \sum_{m=i+1}^{j} \sum_P$).
The summations $[(\sum_{l,m=1}^{i} + \sum_{l=1}^{i}
\sum_{m=i+1}^{j})\sum_P']$ can be calculated  to give
\begin{eqnarray}
 {\cal T}_{ij}^{kink} (i<j,\, j={\frak k})&=&2^{(n-3)}
 (\partial_{\mu} \vec{n}_3) \cdot (\partial_{\nu} \vec{n}_3)
 \,, \\
{\cal T}_{ij}^{kink} (i={\frak k},\,j>i)&=&2^{(n-3)}
(\partial_{\mu} \vec{n}_3) \cdot (\partial_{\nu} \vec{n}_3)\,.
\end{eqnarray}
At the present step of the evaluation, the contribution of the
one-kink configurations starting with $r_1=(-)$ 
to the mixed derivatives takes the form:
\begin{eqnarray}
X^{IJ}_{n(kink)} (r_1\!=\!-)&\!=\!& \left( \frac{g}{2} \right)^n 
\frac{(-2^{(n-3)})}{a^2}
\biggl\{ \sum_{\stackrel{{\frak k}=2}{{\frak k}\, even}}^{n-2} 
\Bigl[
\sum_{i=1}^{{\frak k}-1} \frac{\partial^2 \Pi^{kink}_{-}}{ \partial
q_{i}^{\mu} \partial q_{{\frak k}}^{\nu}} \Bigl|_{q=0} +
\sum_{j= {\frak k}+1}^{n-1} \frac{\partial^2 \Pi^{kink}_{-}}{ \partial
q_{{\frak k}}^{\mu} \partial q_{j}^{\nu}} \Bigl|_{q=0} \Bigr] \biggr\}
\times 
\nonumber
\\
& &\int \! {\rm d}^2x {\rm d}\tau 
(\partial_{\mu}\vec{n}_3) \cdot (\partial_{\nu}
\vec{n}_3)\,.
\label{Xijkinkprelim}
\end{eqnarray}
As to the second derivatives of the product of propagators from
Eq.~(\ref{Pi}), the following cases are encountered:
\begin{eqnarray}
\frac{\partial^2 \Pi^{kink}_-}{\partial q_{i_{even}}^{\mu}
\partial q_{j={\frak k}}^{\nu}} \Bigl|_{q_{i,j}=0}&=&
\frac{\partial^2 \Pi^{kink}_-}{\partial q_{i={\frak k}}^{\mu} \partial
q_{j_{even}}^{\nu}} \Bigl|_{q_{i,j}=0} =
\sum_k \, g_0(k)^{\frac{n}{2}-2}   
\, g_{0,-}(k)^{\frac{{\frak k}}{2}}\,
g_{0,+}(k)^{\frac{n}{2}-\frac{{\frak k}}{2}}\, \nonumber \\
& &\mbox{\hspace*{4.1cm}}\frac{\partial g_0(k-q)}{\partial 
   q^{\mu}}\Bigl|_{q=0}
\,\frac{\partial g_{0}(k-q)}{\partial q^{\nu}} \Bigl|_{q=0}\,,
\nonumber \\
\frac{\partial^2 \Pi^{kink}_-}{\partial q_{i_{odd}}^{\mu} 
\partial q_{j={\frak k}}^{\nu}} \Bigl|_{q_{i,j}=0}&=&
\sum_k \,
g_0(k)^{\frac{n}{2}-1}   \, g_{0,-}(k)^{\frac{{\frak k}}{2}-1} \,
g_{0,+}(k)^{\frac{n}{2}-\frac{{\frak k}}{2}}\, \nonumber \\
& &\mbox{\hspace*{0.6cm}}\frac{\partial g_0(k-q)}{\partial q^{\mu}}
\Bigl|_{q=0}\,\frac{\partial g_{0,-}(k-q)}{\partial q^{\nu}}
\Bigl|_{q=0}\,, \nonumber \\
\frac{\partial^2 \Pi^{kink}}{\partial q_{i={\frak k}}^{\mu}\partial
q_{j_{odd}}^{\nu}} \Bigl|_{q_{i,j}=0} & =&
\sum_k \,g_0(k)^{\frac{n}{2}-1}   \, 
g_{0,-}(k)^{\frac{{\frak k}}{2}} \,
g_{0,+}(k)^{\frac{n}{2}-\frac{{\frak k}}{2}-1}\, \nonumber \\
& &\mbox{\hspace*{0.6cm}}\frac{\partial g_{0}(k-q)}{\partial q^{\mu}}
\Bigl|_{q=0}\,\frac{\partial  g_{0,+}(k-q)}{\partial q^{\nu}}
\Bigl|_{q=0}\,. 
\end{eqnarray}
In the summations over $i$ and $j$ 
of Eq.~(\ref{Xijkinkprelim}), terms with ($i$ even, $j\!=\!{\frak
k}$) or ($i\!=\!{\frak k}$, $j$ even)
appear $(\frac{n}{2}-2)$ times, terms 
with ($i$ odd, $j\!=\!{\frak k}$) appear $\frac{{\frak k}}{2}$ times, 
and terms with ($i\!=\!{\frak k}$, $j$ odd)
appear $(\frac{n}{2}-\frac{{\frak k}}{2})$
times. Accounting for the case $i>j$ and adding the one-kink
configurations starting with $r_1=(+)$, we finally perform
the summation  $\sum_{{\frak k}=2,\,{\frak k}\, even}^{n-2}$
and arrive at the result for the contribution
of the one-kink configurations to the sum $\sum_{r_1 \ldots
r_n}$ in Eq.~(\ref{Xij}) for the mixed derivatives:
\begin{eqnarray}
X^{IJ}_{n(kink)}&\!=\!&\left(\frac{g}{2}\right)^n
\,\frac{2^{(n-2)}}{a^2}\,\Biggl\{\,\sum_k\,g_0(k)^{\frac{n}{2}-2}\,
\frac{\Bigl[g_{0,+}(k)^{\frac{n}{2}-1}
-g_{0,-}(k)^{\frac{n}{2}-1}\Bigr]}{[g_{0,-}(k)-g_{0,+}(k)]}\times
\nonumber \\
& &\mbox{\hspace*{2.8cm}}\Bigl\{(n-4)\,g_{0,-}(k)\,g_{0,+}(k)\,
\frac{\partial  g_0(k-q)}{\partial q^{\mu}}\Bigl|_{q=0}\nonumber \\
& &\mbox{\hspace*{2.8cm}}{}+
\frac{n}{2}\,g_0(k) \frac{\partial [g_{0,-}(k-q)g_{0,+}(k-q)]}{\partial
q^{\nu }} \Bigl|_{q=0} \Bigr\} \nonumber \\
& &{}+ \sum_k\,g_0(k)^{\frac{n}{2}-1}\,
\Bigl[ 1- \frac{g_{0,-}(k)}{g_{0,+}(k)} \Bigr]^{-2}
\Bigl\{ \bigl(\frac{n}{2}-2 \bigr)
\Bigl[g_{0,+}(k)^{\frac{n}{2}}-g_{0,-}(k)^{\frac{n}{2}}\Bigr]
\nonumber \\
& &{}-\frac{n}{2} 
\Bigl[g_{0,-}(k)g_{0,+}(k)^{\frac{n}{2}-1}
-g_{0,+}(k)g_{0,-}(k)^{\frac{n}{2}-1}\Bigr]\Bigr\}\times \nonumber\\
& &\frac{\partial g_0(k-q)}{\partial q^{\mu}} \Bigl|_{q=0}
\frac{\partial [g_{0,+}(k-q)^{-1}g_{0,-}(k-q)]}{\partial q^{\nu}}
\Bigl|_{q=0} \,\Biggr\}\; \int\! dx (\partial_{\mu} \vec{n}_3)\! \cdot\!
(\partial_{\nu}\vec{n}_3)\,.\label{XnIJkink} 
\end{eqnarray}
What remains to be evaluated is the contribution of the two-kink
configurations to the mixed derivatives Eq.~(\ref{Xij}). 
The first sum in
$\Gamma_{ij}^{\mu \nu}$, $\sum_{l,m=1}^{i}$, gives zero for
configurations where the alternation is broken more than once. 
Thus we are left only
with the sum $\sum_{l,=1}^{i} \sum_{m=i+1}^j$, where we first
take $i<j$. The number of the sites after which the first and
second kink occur are denoted by ${\frak k}$ and ${\frak l}$,
respectively. Note that ${\frak k}$, ${\frak l}$ are either
both even or both odd. 
The sum $\sum_{i,j=1,i \neq j}^{n-1}$ only yields a
nonzero contribution for $i={\frak k} \wedge j={\frak l}$, where
only the term with $l\!=\!l_{max}\!=\!{\frak k} \vee
m\!=\!m_{max}\!=\!{\frak l}$ contributes to
$[\sum_{l,=1}^{i}\sum_{m=i+1}^j\sum_P']$. Thus the summations
$[(\sum_{l,m=1}^{i}+ \sum_{l,=1}^{i} \sum_{m=i+1}^j)\sum_P']$ yield
\begin{equation}
{\cal T}_{ij}^{two-kink}=-2^{n-3}\, (\partial_{\mu}
\vec{n}_3) \cdot (\partial_{\nu} \vec{n}_3)\,.
\end{equation}
Upon the determination of the 
second derivatives of the function $\Pi$ for a general
two-kink configuration, it turns out that after adding the 
configurations starting with $r_1=(-)$ and $r_1=(+)$, the form
of $\partial^2 \Pi^{two-kink}/\partial
q_{i={\frak k}}^{\mu}\partial q_{j={\frak l}}^{\nu}$ depends only on
the distance between the kinks. Introducing
${\frak d}={\frak l}-{\frak k}$, we have
\begin{eqnarray}
\frac{\partial^2 \Pi^{two-kink}}{\partial q_{i={\frak k}}^{\mu}
\partial q_{j={\frak l}}^{\nu}}
\Bigl|_{q_{i,j}=0}&=& 
\sum_k \, g_0(k)^{\frac{n}{2}-2}\,
\Bigl[ g_{0,-}(k)^{\frac{n}{2}- \frac{\frak d}{2}} \,
g_{0,+}(k)^{\frac{{\frak d}}{2}}  
+g_{0,+}(k)^{\frac{n}{2}-\frac{\frak d}{2}}\,
g_{0,-}(k)^{\frac{{\frak d}}{2}}
\Bigr] \nonumber\\
& &\frac{\partial g_0(k-q)}{\partial q^{\mu}}\Bigl|_{q=0}
\,\frac{\partial g_{0}(k-q)}{\partial q^{\nu}} \Bigl|_{q=0}\,,
\end{eqnarray}
Now we carry out the sum over all possible two-kink
configurations by summing over all possible distances ${\frak
d}$ (${\frak d}$ runs from 2 to $(n-2)$ over all even values) and
weighting each distance ${\frak d}$
by a factor $(n-{\frak d}-1)$, which is
the number of different two-kink configurations with the same 
inter-kink distance. 
We find the following
contribution of the two-kink configurations to the sum
$\sum_{r_1 \ldots r_n}$:
\begin{eqnarray}
X_{n(two-kink)}^{IJ}&=&\left(\frac{g}{2}\right)^n\,
\frac{2^{(n-2)}}{a^2}
\;\Biggl\{\,\sum_k\,g_0(k)^{\frac{n}{2}-2}\,
\frac{\Bigl[g_{0,+}(k)^{\frac{n}{2}-1}
-g_{0,-}(k)^{\frac{n}{2}-1}\Bigr]}{\Bigl[
g_{0,-}(k)^{-1}-g_{0,+}(k)^{-1}\Bigr]} \nonumber \\
& & \frac{\partial g_0(k-q)}{\partial q^{\mu}}\Bigl|_{q=0}
\,\frac{\partial g_{0}(k-q)}{\partial q^{\nu}} \Bigl|_{q=0}
\Biggl\}\, (n-2) \int \!{\rm d}^2x {\rm d}\tau 
\,(\partial_{\mu} \vec{n}_3)\! \cdot\!
(\partial_{\nu} \vec{n}_3). \label{XnIJtwokink}
\end{eqnarray}
In order to obtain a closed
expression for the fermion determinant, we still have to re-sum
the power series in $n$ from Eq.~(\ref{logseries}) to infinite
order, namely
\bea
\sum_{n=1}^{\infty} \frac{1}{n} \mbox{Tr} \bigl( G_0
\Sigma^{(0)} \bigr)^n=
\sum_{n=1}^{\infty} \frac{1}{n}\Bigl[ X_{n(alt)}^{II} +
X_{n(kink)}^{II} +X_{n(alt)}^{IJ} + X_{n(kink)}^{IJ}
+X_{n(two-kink)}^{IJ} \Bigr]\,, \nonumber \\
& &
\eea
where the terms  on the right-hand side
were defined in Eqs.~(\ref{XnIIalt}),
(\ref{XnIIkink}), (\ref{XnIJalt}), (\ref{XnIJkink}),
and (\ref{XnIJtwokink}).
The summation is achieved in terms of
the polylogarithmic functions \cite{polyl}
$\mbox{Li}_0$, $\mbox{Li}_1$, and $\mbox{Li}_2$, 
which have the power series
expansions
$\mbox{Li}_0(z) = \sum_{n=1}^{\infty}\, z^n$,
$\mbox{Li}_1(z) = \sum_{n=1}^{\infty}\, \frac{z^n}{n}\, =\,
\mbox{ln} \frac{1}{1-z}$, and
$\mbox{Li}_2(z) = \sum_{n=1}^{\infty}\, \frac{z^n}{n^2}$.
 From the integral representation 
of $\mbox{Li}_2(z)$ it follows, for example, that $\frac{d}{dz}
\mbox{Li}_2(z)= -\frac{ \mbox{ln}(1-z)}{z}$. This property will
be needed when the derivatives of these functions, as shown in
Eqs.~(\ref{2ndres})--(\ref{2ndresabb2}) below, are to be
performed for a later numerical evaluation.
In addition, we use the relation
$\sum_{n=1}^{\infty} \, n\, z^n = \frac{z}{(1-z)^2}$.
After some more algebra, we obtain for the
contribution of the second-order derivatives
to the helical part, 
Eq.~(\ref{logseries}), of the fermion determinant: 
\begin{eqnarray}
-\!\!\sum_{\stackrel {n=2}{n even}}^{\infty} 
\frac{1}{n} \left(X_n^{II}+X_n^{IJ}\right)&=& 
\chi^{\mu \nu} \int \!{\rm d}^2x {\rm d}\tau \sum_{a=1}^2
\partial_{\mu}\vec{n}_a\!\cdot \partial_{\nu}\vec{n}_a 
+\tilde{\chi}^{\mu \nu} \int \! {\rm d}^2x {\rm d}\tau\,
\partial_{\mu}\vec{n}_3\cdot\partial_{\nu}\vec{n}_3\,,
\label{2ndres} \nonumber \\
& &
\end{eqnarray}
where we have introduced the abbreviations 
\begin{eqnarray}
\chi^{\mu \nu}&=& \frac{1}{8a^2}\, 
\frac{\partial^2}{\partial q^{\mu} \partial q^{\nu}}
\Bigl[\Phi^-_1(q)+\Phi^+_1(q)\Bigr]\Bigr|_{q=0}\label{2ndresabb1}  \\
\tilde{\chi}^{\mu \nu}&=& \frac{1}{8a^2}\,\frac{\partial^2}
{\partial q^{\mu}  \partial q^{\nu}} \Bigl[ \Phi^-_2(q) +
\Phi^+_2(q) - \Phi^-_3(q) - \Phi^+_3(q)\Bigr]\Bigr|_{q=0}
\nonumber  \\
& &{}+\frac{1}{4a^2}\Bigl[(\Phi^-_4)^{\mu \nu} +(\Phi^+_4)^{\mu
\nu} - (\Phi^-_5)^{\mu \nu} - (\Phi_5^+)^{\mu \nu}
-(\Phi^-_6)^{\mu \nu} - (\Phi_6^+)^{\mu \nu} \Bigr] \,.
\label{2ndresabb2}
\end{eqnarray}
Here $\Phi_1^{\pm}(q)$ was defined in Eq.~(\ref{Phi}), and the
explicit expressions for the functions $\Phi^{\pm}_2(q)$,
$\Phi^{\pm}_3(q)$, $(\Phi^{\pm}_4)^{\mu \nu}$,
$(\Phi^{\pm}_5)^{\mu \nu}$,
 $(\Phi^{\pm}_6)^{\mu \nu}$ are listed in the Appendix.
As can be seen from Eq.~(\ref{Phi}) and 
Eqs.~(\ref{Phi2})-(\ref{Phi6}), 
the parameters of the initial model enter in a
nonperturbative way into the functions $\Phi$, which 
describe the influence of the doping.
\subsection{Ferromagnetic contribution to the fermion determinant}
We are left with the task of evaluating the part of the fermionic
determinant which contains the ferromagnetic  fluctuations
$\vec{L}$, namely the second trace
on the right-hand side of Eq.~(\ref{ferdet}). 
Again we expand up to second order in powers of
$a$.
Whereas in the spiral contribution $\mbox{Tr} \ln(
G_0^{-1}-\Sigma^{(0)})$ every order in the expansion of the
logarithm had to be kept, the ferromagnetic
contribution directly generates an expansion in powers of $a$: 
\be
\label{sferro}
\mbox{Tr} \ln
\bigl[1-a\tilde{G}_0\Sigma^{(1)}-a^2\tilde{G}_0\Sigma^{(2)}\bigr]=
\mbox{Tr} \,\bigl[ -a\tilde{G}_0\Sigma^{(1)}-a^2\tilde{G}_0\Sigma^{(2)}
-\frac{a^2}{2} \tilde{G}_0\Sigma^{(1)}\tilde{G}_0\Sigma^{(1)}
\bigr] .
\ee
In order to evaluate this trace, we must determine the inverse of
$\tilde{G}_0^{-1}=G_0^{-1}-\Sigma^{(0)}$. 
While $G_0^{-1}$ is diagonal in spin and momentum indices
(see Eq.~(\ref{G0})),
$\Sigma^{(0)}$ splits into two parts depending on the wave
vectors $(k-k'-k_s)$ and $(k-k'+k_s)$,
respectively (see Eq.~(\ref{Sigma0})). Since we consider
long-wavelength spin fields, 
$(k-k'\pm k_s)$ must be small. Thus, 
$\Sigma^{(0)}$ has matrix elements only
on four side diagonals between $k$ and $k \pm k_s$.
Hence $\tilde{G}_0^{-1}$ has the form of a band matrix,
where the number of matrix elements is determined by the number
of $k$ points and every $k$ point splits into a $2 \times 2$ 
spin space. Obviously, it is impossible to invert
$\tilde{G}_0^{-1}$ exactly. However, $\tilde{G}_0$ can be found
perturbatively in powers of $a$ by finding a matrix $\bar{G}_0$
which, when multiplied on $\tilde{G}_0^{-1}$, gives a diagonal matrix
$\cal D$ plus off-diagonal terms of $\cal O$$(a)$:
\begin{equation}
\tilde{G}_0^{-1}\,\bar{G}_0={\cal D} +{\cal O}(a) \,.\label{G0G0Q}
\end{equation}
This works because the square of the matrix $\Sigma^{(0)}$
is proportional to the unit matrix
$( \Sigma^{(0)})_{\alpha k, \alpha' k'}^2=g^2
\delta_{\alpha \alpha'} \delta_{k k'}$.
Based on Eq.~(\ref{Sigma0}), we define 
$\bigl(\Sigma^{(0)}_{\pm}\bigr)_{\alpha k, \alpha' k'}=
\frac{g}{2}  \sigma^a n^a_{\pm}(k-k' \pm k_s)$, so that
$\Sigma^{(0)}= \Sigma^{(0)}_- +\Sigma^{(0)}_+$.
Now we try as an ansatz for $\bar{G}_0$:
\begin{equation}
\bar{G}_0= P_1 + P_2  +\Sigma^{(0)}\,, \label{pertdiag}
\end{equation}
multiply it onto $\tilde{G}_0^{-1}$ from the right
and then choose $P_1$ and $P_2$
so that Eq.~(\ref{G0G0Q}) can be fulfilled.
An appropriate choice of $P_1$, $P_2$ should make terms which are 
manifestly non-diagonal in $k$-space either zero or
small and of $\cal O$$(a)$.
Moreover, terms which were already diagonal in $k$-space should
not
lose that property through  multiplication by
$P_{1,2}$. This is accomplished by choosing
\begin{eqnarray}
P_1&=& \Sigma^{(0)}_-\,\Sigma^{(0)}_+\, G_{0,-}^{-1}\, / g^2
\nonumber
\\
P_2&=& \Sigma^{(0)}_+\, \Sigma^{(0)}_-\, G_{0,+}^{-1}\, / g^2
\,,
\end{eqnarray}
where $G_{0,-}^{-1}$ and $G_{0,+}^{-1}$ are exactly diagonal in
spin and momentum space. Their $k$-dependence is chosen to be
\begin{equation}
\bigl(G_{0,\pm}^{-1}\bigr)_{\alpha k,\alpha' k'}
= g_{0,\pm}^{-1} (k)\, \delta_{\alpha\alpha'}
\delta_{kk'} \,.
\end{equation}
Then from the multiplication 
\begin{equation}
\bigl( G_0^{-1}-\Sigma^{(0)}_--\Sigma^{(0)}_+\bigr)
\bigl(P_1 + P_2  +\Sigma^{(0)}_-+\Sigma^{(0)}_+\bigr)=
\mbox{D}-\mbox{F}+ K_- + K_+
\label{DF}
\end{equation}
we obtain a purely diagonal contribution
\begin{equation}
\left(\mbox{D}\right)_{\alpha k,\alpha' k'}=\Bigl[
\frac{1}{2}\,g_{0}^{-1}(k)\, \varrho_{+}(k) -g^2 \Bigr]\,
\delta_{\alpha\alpha'} \delta_{kk'}\,,
\end{equation}
a contribution being approximately diagonal in momentum space
but off-diagonal in spin space
\begin{equation}
\left(\mbox{F}\right)_{\alpha k,\alpha' k'}= \frac{1}{2}
\,g_{0}^{-1}(k)
\, \sigma_{\alpha\alpha'}^a \, n_3^a(k-k') \,\varrho_{-}(k') \,,
\end{equation}
and terms  which are off-diagonal in spin and momentum space,
but which  are of $\cal O$$(a)$:
\begin{eqnarray}
K_{\pm}& =&G_0^{-1} \Sigma^{(0)}_{\pm} - \Sigma^{(0)}_{\pm}
G_{0,\mp}^{-1} \nonumber \\
& =& \bigl[g_0(k) -g_0(k'\mp k_s)\bigr] \sigma_{\alpha
\alpha'}^a
 n_-^a(k-k'\pm k_s) \,.
 \end{eqnarray}
 We have used the abbreviation:
 \begin{equation}
 \varrho_{\pm}(k) =g_{0,-}^{-1}(k) \pm g_{0,+}^{-1}(k)\,.
 \end{equation}
 In order to diagonalize the remaining problem, we multiply Eq.~(\ref{DF})
by a matrix $(\tilde{\mbox{D}}+\tilde{\mbox{F}})$ from the 
right, where
\begin{eqnarray}
(\tilde{\mbox{D}})_{\alpha k,\alpha' k'} &=&\varrho_{-}^{-2}(k)
\Bigl[ \frac{1}{2}
\,g_{0}^{-1}(k)\,\varrho_{+}(k)-g^2  \Bigr] \delta_{\alpha\alpha'}
\delta_{kk'}\,,\\
(\tilde{\mbox{F}})_{\alpha k,\alpha' k'}&=& \frac{1}{2}
\,\varrho_{-}^{-1}(k)\, \sigma_{\alpha\alpha'}^a \, n_3^a(k-k')
\,g_{0}^{-1}(k')\,.
\end{eqnarray}
This leads to the desired expression Eq.~(\ref{G0G0Q}). The matrix
$\bar{G}_0$ which we have been looking for is thus given by
\begin{equation}
\bar{G}_0= \left[ \Sigma^{(0)}_-\, \Sigma^{(0)}_+\,
G_{0,-}^{-1}/g^2
+g^{-2}\Sigma^{(0)}_+\, \Sigma^{(0)}_-\, G_{0,+}^{-1} /g^2
+ \Sigma^{(0)}_- +  \Sigma^{(0)}_+
\right] \left[\tilde{\mbox{D}}+\tilde{\mbox{F}}\right]\,,
\label{barG0}
\end{equation}
and for the diagonal matrix ${\cal D}$ in Eq.~(\ref{G0G0Q}) we obtain
\begin{eqnarray}
({\cal D})_{\alpha k,\alpha' k'}&=& 
\mbox{D}\tilde{\mbox{D}}-\mbox{F}\tilde{\mbox{F}}= \nonumber\\
&=&\varrho_{-}^{-2}(k)\,\Bigl[ 
g^4 - g^2\, g_{0}^{-1}(k)\, \varrho_{+}(k)+g_{0}^{-2}(k)\,  
g_{0,-}^{-1}(k)\,g_{0,+}^{-1}(k)\,\Bigr]
\,\delta_{\alpha\alpha'}\delta_{kk'} \,.
\label{D}
\end{eqnarray}
In order to simplify the notation, we define
\begin{equation}
d(k)=g^4 - g^2\, g_{0}^{-1}(k)\, \varrho_{+}(k)+g_{0}^{-2}(k)\,
g_{0,-}^{-1}(k)\,g_{0,+}^{-1}(k)\,. \label{kleind}
\end{equation}
The small off-diagonal terms are
\begin{equation}
aR = \bigl(\mbox{D}\tilde{\mbox{F}}-
\mbox{F}\tilde{\mbox{D}}) + (K_- +
K_+)(\tilde{\mbox{D}}+\tilde{\mbox{F}})\,.
\end{equation}
 From  Eq.~(\ref{G0G0Q}), 
the inverse of $\tilde{G}_0^{-1}$ to first order in $a$ is then
found to be
\begin{equation}
\tilde{G}_0=\bar{G}_0 {\cal D}^{-1} - a \bar{G}_0 {\cal D}^{-1}R
{\cal D}^{-1}\,.
\end{equation}
This has to be inserted into Eq.~(\ref{sferro}), together 
with the expressions Eqs.~(\ref{Sigma1}), (\ref{Sigma2}) 
for the first and second orders of the fermionic self-energy.
After inserting $\bar{G}_0$ from Eq.~(\ref{barG0}),
$\mbox{Tr} [a\tilde{G}_0\Sigma^{(1)}]$ then becomes a sum of 240 terms,
while $\mbox{Tr}[a^2\tilde{G}_0\Sigma^{(2)}]$ contains 112 terms and
$\mbox{Tr}[\frac{a^2}{2}
\tilde{G}_0\Sigma^{(1)}\tilde{G}_0\Sigma^{(1)}]$ becomes a sum
of 1600 terms. (We  classify different terms by different
combinations of
the order parameter fields. Every individual term still
comprises the momentum- and spin-space summations arising from
the matrix multiplications). 
These numbers grow considerably when the strings
of Pauli matrices contained in these traces are evaluated,
e.g.~a string of six Pauli matrices leads to a sum of 15 different
permutations of contractions.  
While strings of even numbers of Pauli matrices are carried out
using the trace reduction formula Eq.~(\ref{tracered}),
strings of odd numbers of Pauli matrices are evaluated 
by replacing two Pauli matrices  according to the identity
$\sum_{\alpha_2} \sigma^{a_1}_{\alpha_1 \alpha_2}
\sigma^{a_2}_{\alpha_2 \alpha_3}=
\delta^{a_1 a_2} \delta_{\alpha_1  \alpha_3}+i\, 
\varepsilon^{a_1 a_2 a_3} \sigma^{a_3}_{\alpha_1 \alpha_3}$.
Obviously, a string of an odd number of Pauli matrices leads
to inner products and vector products of the involved vectors.
The traces are evaluated with an algebraic program\footnote{The
program {\tt TraceEval} is available via anonymous ftp from
ftp.physik.uni-wuerzburg.de as
/pub/dissertation/klee/TraceEval.m.} which we 
developed using {\em Mathematica} \cite{wolfram}. 
It reduces the trace over the Pauli matrices, 
performs the contractions of the vectors, 
implements the constraint Eq.~(\ref{constr}), and takes the
trace in momentum space by keeping only terms which are
diagonal in the momentum indices. The result for the
ferromagnetic contribution to the fermion determinant 
is 
\begin{eqnarray}
S_{\rm ferro}&=& - \int  \! {\rm d}^2x {\rm d}\tau\, \Bigl\{ 
	      \chi_1\, \vec{L}^2  + \chi_2\, \Bigl[ 
	      (\vec{L}\!\cdot\!\vec{n}_1)^2 + (\vec{L}\!\cdot\! 
	      \vec{n}_2)^2\, \Bigr]
	      + \chi_3\, (\vec{L}\!\cdot\! \vec{n}_3)^2 \nonumber 
	      \\
& &{}+ \frac{\chi_4^{\mu}}{a}
\Bigl[
 (\vec{L}\!\cdot\! \vec{n}_1) ( \vec{n}_1\!\cdot\! \partial_{\mu} 
 \vec{n}_3) +
(\vec{L}\!\cdot\! \vec{n}_2) ( \vec{n}_2\!\cdot\! \partial_{\mu} 
\vec{n}_3) \Bigr]\nonumber \\[0.2cm]
& &{}-\frac{2\chi_4^{\mu}}{a}\,\vec{L}\!\cdot\! \partial_{\mu}\vec{n}_3 
+\frac{\chi_5^{\mu}}{a}\,
(\vec{L}\! \cdot\! \vec{n}_3) ( \vec{n}_1\!\cdot\!
\partial_{\mu} \vec{n}_2) \nonumber \\[0.2cm]
& &{}- \frac{\chi_6^{\mu}}{a}\,\Bigl[ 
(\vec{L}\!\cdot\! \vec{n}_1) ( \vec{n}_2
\!\cdot \!\partial_{\mu} \vec{n}_3) - (\vec{L}\!\cdot\! \vec{n}_2)  
(\vec{n}_1\!\cdot\! \partial_{\mu}
\vec{n}_3) \Bigr]  \nonumber
\\[0.2cm]
& &{}+ \frac{\chi_7^{\mu}}{a}\, \vec{L}\!\cdot\!  \Bigl( 
\vec{n}_1 \times \partial_{\mu} \vec{n}_1 +\vec{n}_2\times 
\partial_{\mu} \vec{n}_2 \Bigr) +
\frac{\chi_8^{\mu}}{a}\, \vec{L}\!\cdot\! \Bigl(\vec{n}_3\times
\partial_{\mu} \vec{n}_3\Bigr) \Bigr] \Bigr\} \,.
\label{Sferro}
\end{eqnarray}
The quantities $\chi_1 \ldots \chi_3, \chi_4^{\mu} \ldots \chi_8^{\mu}$ 
are generalized susceptibilities of
the fermions in the presence of the spin fields. The explicit 
expressions are listed in the Appendix. 
The functions $\chi$ consist of summations over 
$\epsilon_n$ and $\vec{k}$ of polynomials of the 
inverse free fermionic propagator
$g_0^{-1}$ and its derivative. 
Shifting the summation index $\vec{k} \rightarrow -\vec{k}$
and using the symmetry of the free propagator,
 Eq.~(\ref{g0sym1}), it can be shown that 
\be
\begin{array}{lll}
\chi_4^{\mu} = 0 &\quad  \mbox{for}\quad   &\mu=0\,, \\
\chi_5^{\mu} =\chi_6^{\mu}=\chi_7^{\mu}=\chi_8^{\mu}= 0 & \quad 
\mbox{for}\quad&  \mu=1,2\,,
\label{suszcomp}
\end{array}
\ee
i.e.~$\chi_4^{\mu}$ is non-zero
only for the spatial components $\mu=1,2$, while $\chi_5^{\mu}$,
$\chi_6^{\mu}$, $\chi_7^{\mu}$, $\chi_8^{\mu}$ are non-zero
only for the time-component $\mu=0$.
Note that the
susceptibilities are proportional to $d^{-1}(k)$ or $d^{-2}(k)$,
where $d(k)$ was defined in Eq.~(\ref{kleind}). The zeroes of $d(k)$
determine the dispersion relations of the holes in the 
helical spin background. Since $d(k)$ is 
fourth order in the free inverse propagators $g_0^{-1}$, which
contain contributions from two bands, $d(k)$ is eighth order
in $i \epsilon_n$, and eight hole bands will result. 
This additional splitting in comparison to the
antiferromagnetic case, in which four hole bands were obtained
\cite{muze}, results from the fact that the momentum transfer on the
fermions interacting with the spin field is now  $Q \pm\delta Q$ instead
of $Q$ as in the antiferromagnetic case.
\subsection{Contribution of the Heisenberg part}
Finally we have to calculate the continuum limit of
the pure spin part of the action, Eq.~(\ref{Sspin}). 
The study of the continuum limit of this  action, 
to which our model simplifies in the absence of
doping, has in itself attracted  much interest.
In order to investigate the effect of frustration on the
stability of the ordered state against quantum fluctuations,
field-theoretic mappings have been performed for
the Heisenberg antiferromagnet on a triangular lattice
\cite{domb,antimotri,apel,ADM93}, for frustrated
Heisenberg antiferromagnets 
on a chain and on $d$-dimensional lattices 
\cite{antimo_1D_dD,senec}.
The frustrating effect of either the lattice geometry or the 
competition between nearest-neighbour and higher-order
interactions leads to noncollinear classical ground states, so
that the order parameter is an element of SO(3).
%
%
Thus our gradient expansion of the pure spin part, carried
out in position space, is similar to the above-mentioned treatments.
However, in these works
the spins in the ground state 
have a periodicity, so that one can deal with a finite number
of sublattices. This does not apply in 
the case of the incommensurate spiral.

It is useful to extract the antiferromagnetic modulation from
$\vec{n}$
by writing $\vec{n}_{pq}=(-1)^{(p+q)}\,\vec{m}_{pq}$, where we
define
\begin{equation}
\vec{m}_{pq}= \vec{n}_1 \cos (\delta \vec{Q}\! \cdot\! \vec{r}_{pq})
- \vec{n}_2 \sin (\delta \vec{Q} \!\cdot \!\vec{r}_{pq})
\end{equation}
First, we consider the Berry phase.
We inject Eq.~(\ref{ansexp}) for $\vec{S}_{pq}$ and expand 
$\vec{A}(\vec{S}_{pq}/S)$ and 
$\partial_{\tau}\vec{S}_{pq}$ to first order in $a$.
We exploit the gauge freedom that we have in the definition
of $\vec{A}$ to choose it an even or an odd function of its
argument, use the constraint satisfied by the monopole potential,
Eq.~(\ref{monpolpot}), and the fact that 
total time derivatives drop out upon $\tau$-integration.
Discarding terms which are of third or higher order in
derivatives of $\vec{m}$, we obtain:
\begin{equation}
S_{\rm Berry}=\int_0^{\beta}{\rm d}\tau \,iS\sum_{pq} \left[
(-1)^{(p+q)}\vec{A}(\vec{m}_{pq}) \cdot\partial_{\tau}\vec{m}_{pq}
+a\, \vec{L} \cdot( \vec{m}_{pq} \times \partial_{\tau}\vec{m}_{pq})
\right]\,. \label{sberry}
\end{equation}
 From the term involving the vector potential in
Eq.~(\ref{sberry}), we first consider one 
chain of lattice sites by holding the
index $q$ fixed:
\begin{eqnarray}
S_{\vec{\scriptstyle A}}(q)&=&
\lefteqn{\int_0^{\beta}{\rm d}\tau\, iS \sum_p (-1)^p \vec{A}(\vec{m}_{pq})
 \cdot \partial_{\tau}\vec{m}_{pq} =} \nonumber\\
&= &\int_0^{\beta}{\rm d}\tau\, iS
\sum_{p_{even}} \Bigl[ \vec{A}(\vec{m}_{pq})\cdot \partial_{\tau}
\vec{m}_{pq}-
\vec{A}(\vec{m}_{p+1,q})\cdot \partial_{\tau}\vec{m}_{p+1,q}\Bigr]\,.
\label{SAq}
\end{eqnarray}
Expanding our smooth order parameter fields according to
$\vec{n}_{1,2}(p+1,q)=
\vec{n}_{1,2}(p,q) +a \partial_x \vec{n}_{1,2}(p,q) $, we find:
\begin{eqnarray}
\vec{m}_{p+1,q}&=& \vec{m}_{pq} \cos (\delta Q_x a)-
		   \vec{{\tilde m}}_{pq}  \sin (\delta Q_x a) 
		   \nonumber \\
&  &+ a\,(\Delta_x \vec{m}_{pq}) \cos (\delta Q_x a)-
	  a\,(\Delta_x \vec{{\tilde m}}_{pq})  \sin (\delta Q_x a)
	   \,, \label{mp+1}
\end{eqnarray}
where we defined
$\vec{{\tilde m}}_{pq}= \vec{n}_1 \sin (\delta \vec{Q} \!\cdot
\!\vec{r}_{pq})
+ \vec{n}_2 \cos (\delta \vec{Q} \!\cdot \!\vec{r}_{pq})$.
Note that in contrast
to collinear spin configurations, the zeroth order
contribution to $\vec{m}_{p+1,q}$ is not 
$\vec{m}_{pq}$, but a vector which we denote by
$\vec{m}_{p+1,q}^0= \vec{m}_{pq} \cos (\delta Q_x
a)-\vec{{\tilde m}}_{pq}  \sin (\delta Q_x a)$. With expressions like
$\Delta_x \vec{m}_{pq}$  we mean:
\begin{equation}
\Delta_x \vec{m}_{pq}= (\partial_x\vec{n}_1) \cos (\delta \vec{Q}
\!\cdot \!\vec{r}_{pq})-(\partial_x\vec{n}_2) \sin (\delta \vec{Q}
\!\cdot \!\vec{r}_{pq})\,.
\end{equation}
After some algebra one obtains
\begin{eqnarray}
\vec{A}(\vec{m}_{p+1,q})\cdot \partial_{\tau}\vec{m}_{p+1,q}&=&
\vec{A}(\vec{m}_{p+1,q}^0)\cdot \partial_{\tau} \vec{m}_{p+1,q}^0
\nonumber \\ 
& & - a \,\vec{m}_{p+1,q}^0 \cdot (
\Delta_x \vec{m}_{p+1,q}^0  \times \Delta_{\tau}
\vec{m}_{p+1,q}^0)\,. \label{Amp+1}
\end{eqnarray}
Now,  the terms
$\int_0^{\beta}{\rm d}\tau  \vec{A}(\vec{m}_{pq})\cdot \partial_{\tau}
\vec{m}_{pq}$ and $\int_0^{\beta}{\rm d}\tau
\vec{A}(\vec{m}_{p+1,q}^0)\cdot \partial_{\tau} \vec{m}_{p+1,q}^0$
describe the area of the caps bounded by the paths of
$\vec{m}_{pq}$
and $ \vec{m}_{p+1,q}^0$, respectively.  Since $\vec{m}_{pq}$
and $\vec{m}_{p+1,q}^0$  are connected by a space-time 
independent rotation 
about the axis $\vec{n}_3$ by the angle $(\delta Q_x a)$, the solid
angles spanned by their paths are equal. Thus the corresponding
terms cancel when (\ref{Amp+1}) is inserted into Eq.~(\ref{SAq}).
After some tedious 
algebra, we get for the
remaining term:
\begin{eqnarray}
\vec{m}_{p+1,q}^0 \cdot ( \Delta_x \vec{m}_{p+1,q}^0  \times
\Delta_{\tau}  \vec{m}_{p+1,q}^0)&= &- \vec{n}_1\! \cdot\! (
\partial_{\tau} \vec{n}_1 \times  \partial_x \vec{n}_1)
\cos (\delta \vec{Q}\!\cdot\! \vec{r}_{pq} + \delta Q_x )
\nonumber\\
& &+ \vec{n}_2\! \cdot\! ( \partial_{\tau} \vec{n}_2\times  \partial_x
\vec{n}_2) \sin (\delta \vec{Q}\!\cdot\! \vec{r}_{pq} + \delta Q_x )
\end{eqnarray}
Employing the relations $\epsilon^{abc}  n_d^a\, \partial_{\tau}
n_d^b\, \partial_x n_d^c = \epsilon_{def} \partial_{\tau} n_e^a
\,\partial_x n_f^a$, this
result can be shown to agree with the one obtained 
for a spiral configuration 
in a frustrated Heisenberg chain within a different method
\cite{antimo_1D_dD}. Due to dimensional reasons, we must have: $\delta
\vec{Q} = \delta \vec{\varphi}/ a$. We assume that in the continuum
limit the angle $\delta \vec{\varphi}$ between neighbouring spins
remains constant. Thus in the limit $a \rightarrow 0$ the
terms containing 
$ \cos (\delta \vec{\varphi}\cdot \vec{x} /a)$, $ \sin (\delta
\vec{\varphi}\cdot \vec{x} /a)$ oscillate very rapidly and drop
out. Then the total Berry phase is given by
the second term in Eq.~(\ref{sberry}), from which we keep only
non-oscillatory contributions. Taking the  continuum limit
$\lim_{\,a \rightarrow 0} \sum_{pq}\,a^2=\int {\rm d}^2 x$ 
leads to
\begin{equation}
S_{\rm Berry}=\int_0^{\beta}{\rm d}\tau \int {\rm d}^2x 
\,\,  \frac{iS}{2a} \,
\vec{L}\cdot \bigl( \vec{n}_1\times \partial_{\tau} \vec{n}_1 +
\vec{n}_2 \times \partial_{\tau} \vec{n}_2 \bigr) \;.
 \label{Sberry}
\end{equation}
We now turn to the 
real term in Eq.~(\ref{Sspin}).
We attribute to each site the quantity 
\begin{equation}
\frac{1}{2}\;J_H \;\Biggl\{ \sum_{\delta=\pm 1}
\vec{S}_{pq}\cdot\vec{S}_{p+\delta,q}+\sum_{\delta'=\pm 1}
\vec{S}_{pq}\cdot \vec{S}_{p,q+\delta'} \Biggr\}
\end{equation}
and then perform the sum over all sites. 
As before,  we insert Eq.~(\ref{ansexp}) for $\vec{S}$ and expand 
the fields at site $(p+\delta,q)$ in
terms of the corresponding fields at site $(p,q)$ using 
Eq.~(\ref{mp+1}). Again, we eliminate short-range oscillatory
contributions. After some  lengthy but straightforward
algebra we obtain 
\begin{eqnarray}
S_{\rm real}&=&-\int_0^{\beta} \! {\rm d}\tau \int {\rm d}^2x 
\,\, J_{\rm H}\,
	    S^2\, \Biggl\{\, \sum_{\mu=1}^2
	    a^{-1} \sin (\delta Q_{\mu}a)\, 
	    \vec{n}_1\cdot \partial_{\mu} \vec{n}_2 \nonumber \\
& &+ \sum_{\mu=1}^2 \frac{1}{4} \cos(\delta Q_{\mu}a)  \Bigl[ 
    (\partial_{\mu} \vec{n}_1)^2 + (\partial_{\mu} \vec{n}_2)^2 \Bigr]
+ \Bigl[\, 2+  \sum_{\mu=1}^2 \cos(\delta Q_{\mu}a)\Bigr]
     \, \vec{L}^2 \nonumber \\
& &- \Bigl[\, 2+ \frac{1}{2}\sum_{\mu=1}^2 \cos(\delta Q_{\mu}a)
     - \frac{1}{2} \sum_{\mu=1}^2 \cos^2(\delta Q_{\mu}a)
    \Bigr]   \Bigl[ (\vec{L}\cdot \vec{n}_1)^2 +(\vec{L} \cdot
    \vec{n}_2)^2\Bigr] \Biggr\} \,.
    \nonumber \\
    & &
\label{Sreal}
\end{eqnarray}
Note that here the index $\mu$ runs only over the space
components, namely $\mu=1,2$. 
\section{Discussion of the continuum theory}
In order to express our continuum action in terms of the
SO(3) order parameter $\hat{Q}=(\vec{n}_1\, \vec{n}_2\, \vec{n}_3)$,
the ferromagnetic fluctuations $\vec{L}$
have to be integrated out. Since our resulting action is 
linear and quadratic in $\vec{L}$, this can be done by
extremizing the action with respect to $\vec{L}$.
Inserting Eqs.~(\ref{Sferro}), (\ref{Sberry}),
and (\ref{Sreal}) into the saddle-point equation
$\frac{\delta}{\delta  L^a} \bigl( 
S_{\rm ferro}+S_{\rm Berry}+S_{\rm real} \bigr)   = 0$,
we obtain
\begin{equation}
\left[ \bar{\chi}_1\, \delta^{ab}+ \bar{\chi}_2\, \left(n_1^a n_1^b 
+n_2^a n_2^b \right) + \chi_3\, n_3^a n_3^b \right]\,L^b=v^a\,,
\label{saddle}
\end{equation}
\nopagebreak
where $\vec{v}$ denotes the sum of all terms which were
multiplied by  $\vec{L}$ linearly in the action
\bea
v^a&=& \frac{1}{2a} \Bigl[ \,-2 \chi_4^{\mu}\, \partial_{\mu}
n_3^a +\chi_4^{\mu} \sum_{i=1}^2 n_i^a\,n_i^b
\partial_{\mu}n_3^b 
-\chi_5\, n_3^a\, \epsilon^{bcd}\, n_3^b\, n_2^c \partial_{\mu} n_2^d
\nonumber \\[-0.1cm]
& &{} +\chi_6^{\mu}\sum_{{i,j=1 \atop i \neq j}}^2 
n_i^a\, \epsilon^{bcd}
\,n_i^b \,n_j^c \partial_{\mu} n_j^d 
+\bar{\chi}_7^{\mu}\, \epsilon^{abc}\sum_{i=1}^2 n_i^b
\partial_{\mu}n_i^c +\chi_8^{\mu}\,
\epsilon^{abc}n_3^b\partial_{\mu}n_3^c\Bigr]\,,
\eea
and we have introduced the following abbreviations to
incorporate fermionic and Heisenberg contributions:
\begin{eqnarray}
\bar{\chi}_1& =& \chi_1 - J_{\rm H} S^2 
	       \Bigl[ 2+  {\textstyle \sum_{\mu=1}^2}
	    \cos(\delta Q_{\mu}a)\Bigr]\,, \nonumber \\
\bar{\chi}_2&=& \chi_2 + J_{\rm H} S^2 \Bigl[ 2+ \frac{1}{2}
	      {\textstyle \sum_{\mu=1}^2} \cos(\delta Q_{\mu}a) -
	      \frac{1}{2} {\textstyle \sum_{\mu=1}^2} 
	      \cos^2(\delta Q_{\mu}a) \Bigr] \,, \nonumber \\
\bar{\chi}_7^{\mu=0}&=&\chi_7^{\mu=0}-\frac{iS}{2} \,. \label{barchis}
 \end{eqnarray}
We solve the saddle-point equation for $\vec{L}$ by multiplying
both sides of Eq.~(\ref{saddle}) by the inverse of the matrix
$ \left[ \bar{\chi}_1\, \delta^{ab}+ \bar{\chi}_2\, \left(n_1^a
n_1^b +n_2^a n_2^b \right) + \chi_3\, n_3^a n_3^b \right]$ 
and obtain
\begin{eqnarray}
aL^a&=&{} - \zeta_1\, \bar{\chi}_7^{o}\, \epsilon^{abc}\,  (\,n_1^b
\,\partial_{\tau}n_1^c+ n_2^b \,\partial_{\tau}n_2^c )
- \zeta_1\, \chi_8^{o}\, \epsilon^{abc}\, n_3^b
\,\partial_{\tau}n_3^c 
\nonumber \\
& &{}-n_1^a
\bigl[ \, \zeta_2^{\alpha}\, n_1^b \,\partial_{\alpha}n_3^b - 
\zeta_3^{o}\,
n_2^b \,\partial_{\tau}n_3^b \bigr] \nonumber \\
& &{}-n_2^a \bigl[\, \zeta_2^{\alpha}
\, n_2^b \,\partial_{\alpha}n_3^b +\zeta_3^{o}\, n_1^b\, 
\partial_{\tau}n_3^b \bigr]  \nonumber \\
& &{}+ \zeta_4^{o} \,n_3^a\, n_1^b\, \partial_{\tau} n_2^b 
 + 2\, \zeta_1\,
\chi_4^{\alpha}\, \partial_{\alpha}n_3^a \,. \label{L}
\end{eqnarray}
The newly introduced abbreviations $\zeta_1$, $\zeta_2^{\mu}$,  
$\zeta_3^{\mu}$, and $\zeta_4^{\mu}$ 
are defined in the Appendix.
Recall that the upper index $0$ on the generalized fermionic
susceptibilities 
denotes their frequency component, which accompanies the time 
component of the field
derivatives, while the index $\alpha=1,2$ runs over the
wave-vector components of the susceptibilities, which accompany 
the spatial
components of the field derivatives. In the above solution for
$\vec{L}$, Eq.~(\ref{L}), we have applied the results of 
Eq.~(\ref{suszcomp}) to show explicitly how $\vec{L}$ is composed
of first-order space and time derivatives of $\hat{Q}
=(\vec{n}_1\, \vec{n}_2\, \vec{n}_3)$.
Our expression for $\vec{L}$ must now
be reinserted into Eqs.~(\ref{Sferro}), (\ref{Sberry}), 
(\ref{Sreal}) for $S_{\rm ferro}$, $S_{\rm Berry}$, and
$S_{\rm real}$.
In order to simplify the result, the following identities,
derived from the constraint on the order parameter,
Eq.~(\ref{constr}), are needed:
\bea
(\vec{n}_1\!\cdot\! \partial_{\mu}\vec{n}_3)(\vec{n}_1\!\cdot\!
\partial_{\nu}\vec{n}_3)+(\vec{n}_2\!\cdot\!
\partial_{\mu}\vec{n}_3)(\vec{n}_2
\!\cdot\!\partial_{\nu}\vec{n}_3)& =&
\partial_{\mu}\vec{n}_3\cdot\partial_{\nu}\vec{n}_3\,,
\nonumber \\
(\vec{n}_1\!\cdot\!
\partial_{\mu}\vec{n}_3)(\vec{n}_2\!\cdot\!\partial_{\nu}\vec{n}_3)-
(\vec{n}_1\!\cdot\!
\partial_{\nu}\vec{n}_3)(\vec{n}_2\!\cdot\!\partial_{\mu}\vec{n}_3)&=&
\vec{n}_3\cdot \bigl(\partial_{\mu}\vec{n}_3 \times
\partial_{\nu}\vec{n}_3\bigr)\,, \nonumber \\
\partial_{\nu}\vec{n}_3\cdot \bigl(
\vec{n}_1 \times \partial_{\mu}\vec{n}_1 + \vec{n}_2 \times
\partial_{\mu}\vec{n}_2 \bigr)
&=&\vec{n}_3\cdot \bigl(\partial_{\mu}\vec{n}_3 \times
\partial_{\nu}\vec{n}_3\bigr)\,.
\eea
The total $\vec{L}$-dependent part of the action can then be
cast in the form 
\begin{eqnarray}
S_{\vec{\scriptstyle L}}&=&- \bar{\chi}^{\mu \nu}\, \int 
{\rm d}^2x {\rm d}\tau\, \bigl(
\partial_{\mu}\vec{n}_1\cdot \partial_{\nu}\vec{n}_1 +
\partial_{\mu}\vec{n}_2\cdot \partial_{\nu}\vec{n}_2 \bigr)
\nonumber \\
& &{}-\left(\bar{\bar{\chi}}^{\mu \nu}-\bar{\chi}^{\mu \nu}\right)
\, \int {\rm d}^2x {\rm d}\tau\,
\partial_{\mu}\vec{n}_3\cdot\partial_{\nu}\vec{n}_3\nonumber\\
& &{}-\hat{\chi}^{\mu \nu}\,\int {\rm d}^2x {\rm d}\tau\,
\vec{n}_3\cdot\bigl(\partial_{\mu}
\vec{n}_3 \times \partial_{\nu}\vec{n}_3\bigr)\,, \label{Lres}
\end{eqnarray}
where the prefactors are given by 
\begin{eqnarray}
\bar{\chi}^{\mu \nu} &=&\frac{1}{2a^2}\left[ \bar{\chi}_1+
\chi_3\right]^{-1}
\left[ \frac{1}{4} \chi_5^{\mu}\chi_5^{\nu} + \bar{\chi}_7^{\mu}
\bar{\chi}_7^{\nu} - \chi_5^{\mu}\bar{\chi}_7^{\nu} 
\right]\,,
\label{barchi}\\
\bar{\bar{\chi}}^{\mu \nu}&=&\frac{1}{4a^2}\left[
\bar{\chi}_1+\bar{\chi}_2\right]^{-1}
\bigl[ \chi_4^{\mu}\chi_4^{\nu} + \chi_6^{\mu}\chi_6^{\nu} + 
\bar{\chi}_7^{\mu}\bar{\chi}_7^{\nu} +
\chi_8^{\mu}\chi_8^{\nu}  \nonumber\\
& &\hspace*{2.8cm}{}+2\chi_6^{\mu}\bar{\chi}_7^{\nu} +
2\chi_6^{\mu} \chi_8^{\nu} +
2\bar{\chi}_7^{\mu}\chi_8^{\nu}\bigr] \,,\label{barbarchi}\\
\hat{\chi}^{\mu \nu}&=& \frac{1}{2a^2}\left[\bar{\chi}_1+\bar{\chi}_2
\right]^{-1}
\bigl[\chi_6^{\mu} +\bar{\chi}_7^{\mu}+\chi_8^{\mu}\bigr]\,
\chi_4^{\nu}\,. \label{hatchi}
\end{eqnarray}
 From Eqs.~(\ref{1stres}), (\ref{2ndres}) --  (\ref{2ndresabb2})
for the spiral contribution to the fermion determinant, 
 Eq.~(\ref{Sreal}) for the contribution of the Heisenberg part,
 and Eq.~(\ref{Lres}) for the contribution of the
 $\vec{L}$-dependent part of the action, we find 
 the total continuum theory to be given by
an SO(3)  quantum nonlinear $\sigma$ model, a term linear in
the derivatives and a geometric term which is third order in
the fields and contains first-order
derivatives with respect to time and space,
\be
S= S_{\rm qnl\sigma m} + S_{\rm lin} + S_{\rm geom}\,.
\ee
The final form of our quantum nonlinear $\sigma$ model action is
\begin{equation}
S_{\rm qnl\sigma m}= - \int \! {\rm d}^2x {\rm d}\tau\, 
		   \sum_{a=1}^3\, p_a^{\mu \nu} \partial_{\mu}
\vec{n}_a(x) \cdot \partial_{\nu}\vec{n}_a(x)\,, \label{act1}
\end{equation}
where $p_1^{\mu \nu}= p_2^{\mu \nu} \neq p_3^{\mu \nu}$. 
The coupling constants of the model are given  as a function of the 
microscopic parameters and the generalized fermionic susceptibilities:
\begin{eqnarray}
p_1^{\mu \nu}\;\;=\;\;\,p_2^{\mu \nu}&=& \bar{\chi}^{\mu \nu}
	    - \chi^{\mu \nu} +\frac{J_{\rm H}
	    S^2}{4}\cos (\delta Q_{\mu}a) 
	    \, \delta^{\mu \nu}\,,\label{coefp1} \\[0.1cm]
p_3^{\mu \nu}&=& \bar{\bar{\chi}}^{\mu \nu}-\bar{\chi}^{\mu \nu} - 
		 \tilde{\chi}^{\mu \nu}\,.\label{coefp3}
\end{eqnarray}
Equivalently to Eq.~(\ref{act1}),
the model can be written in SO(3) matrix form:
\begin{equation} 
S_{\rm qnl\sigma m}= - \int {\rm d}^2x {\rm d}\tau\,  
\mbox{Tr}\, \bigl[ \hat{P}^{\mu \nu}\,
\partial_{\mu}\hat{Q}^T(x)\,  \partial_{\nu}\hat{Q}(x) \bigr]\,,
\label{act2}
\end{equation}
where the coefficient matrix is given by 
\be
\hat{P}^{\mu \nu}=
\left( \begin{array}{ccc}
p_1^{\mu \nu}& 0 & 0 \nonumber  \\ 
0& p_1^{\mu \nu}& 0 \nonumber  \\
0 & 0& p_3^{\mu \nu} 
\end{array} \right) \,. \label{matrP}
\ee
Among the nine fields $Q_{ab}(x)=n^a_b(x)$, taking into account
the constraint Eq.~(\ref{constr}), there are three independent
fluctuating fields. Hence, our
continuum action describes three massless modes, or spin waves,
which result from the complete breaking of the O(3) rotation group
by the ground state. Each spin wave excitation 
corresponds to infinitesimal
rotations about one of the $\vec{n}_a$'s.
One Goldstone mode describes variations of the spin
orientation within the plane defined by the classical
ground state, and thus
corresponds to infinitesimal rotations about the axis 
$\vec{n}_3$, while
the other two Goldstone modes describe out-of-plane fluctuations,
and are related to rotations about $\vec{n}_1$ and $\vec{n}_2$.
Eqs.~(\ref{act1}) or (\ref{act2}) describe the linear spectrum
and the interactions of these spin-wave excitations.

The coefficient matrix $\hat{P}$ in Eq.~(\ref{matrP}) is represented 
in internal space. Every matrix element of $\hat{P}^{\mu \nu}$,
however, is  still a matrix in the space-time indices $\mu$ and
$\nu$.  We will now consider the symmetry properties of the
matrices $p_1^{\mu \nu}$ and $p_3^{\mu \nu}$.
As regards $p_1^{\mu \nu}$, it can be seen from
Eq.~(\ref{coefp1}) that it is
symmetric in $\mu$, $\nu$ when looking at the definitions of 
$\chi^{\mu \nu}$, Eq.~(\ref{2ndresabb1}),
and $\bar{\chi}^{\mu \nu}$, Eq.~(\ref{barchi}), and taking into account
Eq.~(\ref{suszcomp}), which implies that $\bar{\chi}^{\mu \nu}$
is nonzero only for $\mu=\nu=0$.
Inspection of the terms appearing in Eq.~(\ref{coefp3}) for
$p_3^{\mu \nu}$ (see Eqs.~(\ref{2ndresabb2}), (\ref{barchi}) and 
(\ref{barbarchi})),
shows that $p_3^{\mu \nu}$ contains also antisymmetric
contributions. However, since we sum over $\mu$ and $\nu$
in Eq.~(\ref{act1}) or (\ref{act2}),  and the term
$(\partial_{\mu} \vec{n}_a(x) \cdot \partial_{\nu}\vec{n}_a(x))$
is symmetric  in $\mu$, $\nu$, the antisymmetric contributions
cancel and we are left with a symmetric coefficient matrix.

Let us now consider the matrix elements $\hat{P}^{\mu \nu}$
which mix frequency and wave-vector indices (or, in the
corresponding field derivatives, space and time indices).
For the spiral contributions to the generalized fermionic
susceptibilities, it can be shown by performing the derivatives
and using the symmetry of the free propagator, Eq.~(\ref{g0sym1}),
that for $\alpha =1,2$
\bea
\frac{\partial^2}{\partial q^o \partial q^{\alpha}}\, \Phi_{1,2,3}^+
(q) \Bigl|_{q=0}&=&- \frac{\partial^2}{\partial q^o \partial
q^{\alpha}}\, \Phi_{1,2,3}^-(q) \Bigl|_{q=0} \,, \\[0.3cm]
\bigl( \Phi_{4,5,6}^+ \bigr)^{o\alpha}&=&- \bigl(
\Phi_{4,5,6}^-\bigr)^{o\alpha}\,.
\eea
As a consequence, $\chi^{o\alpha}=\chi^{\alpha o}$ 
and $\tilde{\chi}^{o\alpha}=\tilde{\chi}^{\alpha o}$
vanish, as can  be seen from Eqs.~(\ref{2ndresabb1}) and 
(\ref{2ndresabb2}). 
For the ferromagnetic contributions, we know that 
$\bar{\chi}^{\mu \nu} \neq 0$ only for $\mu=\nu=0$ and
that  $\bar{\bar{\chi}}^{\mu \nu}$ does also not
mix frequency and wave-vector indices (see
Eqs.~(\ref{suszcomp}) and (\ref{barbarchi})),  so that
\be
\begin{array}{ccccc}
p_1^{o\alpha}&=&p_1^{\alpha o}& =& 0\,, \\
p_3^{o\alpha}&=&p_3^{\alpha o}& =& 0\,. 
\end{array}
\ee
Therefore, the matrices $p_1^{\mu \nu}$ and
$p_3^{\mu \nu}$  are of block-diagonal form (for clarity, 
we will use  $\tau$ instead of 0 for the frequency component of
the coefficients, and
$\alpha= x,y$ instead of $\alpha= 1,2$ for their wave-vector
components),
\be
\hat{p}_a=
\left( \begin{array}{ccc}
p_a^{\tau \tau}& 0 & 0 \nonumber  \\
0& p_a^{xx} & p_a^{xy} \nonumber  \\
0& p_a^{yx} & p_a^{yy}
\end{array} \right)\,, \;\;\;\;\; a=1,3 \,.\label{blockdia}
\ee
The action, Eq.~(\ref{act1}), can then be written as
\be
S_{\rm qnl\sigma m}= - \int {\rm d}^2x {\rm d}\tau\, \mbox{Tr}\, \Bigl[
\hat{P}^{\tau \tau}\, \partial_{\tau} \hat{Q}^T(x)\,
\partial_{\tau}\hat{Q}(x)+ \sum_{\alpha , \beta
=x,y} \hat{P}^{\alpha \beta} \,
\partial_{\alpha}\hat{Q}^T(x)\,  \partial_{\beta}\hat{Q}(x)
\Bigr]\,.\label{act3}
\ee
This form, in which the coefficient of the time derivative term
is a tensor in spin space and the coefficient of the space
derivative term is a tensor both in spin space and in real space,  
was suggested as the most general case for systems with noncollinear 
spin orientation within a hydrodynamical theory \cite{halpsas}. For some 
special cases of the direction of the spiral wave vector, the matrices 
$\hat{p}_a$ become diagonal. If,  as suggested  by experiments \cite{Cheo} 
on cuprate superconductors and by theoretical studies \cite{shraimansiggia}
of models related to the spin-fermion model,  
the spiral wave vector lies along the (1,0) or (0,1) directions
of the lattice, i.e.~$\vec{k}_s=(\frac{\pi}{a}\pm \delta Q_x, 
\frac{\pi}{a})$ or $\vec{k}_s=(\frac{\pi}{a},\frac{\pi}{a}\pm
\delta Q_y)$, then, using the symmetry relation
 Eq.~(\ref{g0sym2}) it can be shown that
\be
\begin{array}{ccc}
\displaystyle{\frac{\partial^2}{\partial q^x \partial q^y}}\,
\Phi_{1,2,3}^{\pm}(q) \Bigl|_{q=0} &=&0 \,,\nonumber \\[0.3cm]
\bigl( \Phi_{4,5,6}^{\pm} \bigr)^{xy} &=&0 \,,\nonumber \\[0.3cm]
\bar{\bar{\chi}}^{xy}&=&0 \,.\nonumber 
\end{array}
\ee
Thus, Eqs.~(\ref{coefp1}) and (\ref{coefp3}) yield 
\be
\hat{p}_a=
\left( \begin{array}{ccc}
p_a^{\tau \tau}& 0 & 0 \nonumber  \\
0& p_a^{xx} &0\nonumber  \\
0&0& p_a^{yy}
\end{array} \right)\,.
\ee
In the general case of  Eq.~(\ref{blockdia}),
the symmetric $(2 \times 2)$ matrix in $x$ and $y$
can be diagonalized, which  amounts to changing to a new
basis which is a linear combination of the lattice basis
vectors.

In order to make the properties of the spin waves  resulting
from our theory more transparent, we write the action in 
another equivalent form 
\be
S_{\rm qnl\sigma m}= - \int {\rm d}^2x {\rm d}\tau\, \mbox{Tr}\, \bigl[ 
\hat{P}_{\rm diag}^{\mu
\mu}\, \bigl( \hat{Q}^{T} \partial_{\mu}\hat{Q}\bigr)^2\bigr] 
\label{diagact}
\ee
(where we assume the coefficient matrix to be diagonalized).
Now we represent the three degrees of freedom contained in the 
antisymmetric spin-space matrix $( \hat{Q}^{-1}
\partial_{\mu}\hat{Q}\bigr)$  according to $(
\hat{Q}^{-1}\partial_{\mu}\hat{Q}\bigr)_{12}= A_{\mu}$,
$( \hat{Q}^{-1} \partial_{\mu}\hat{Q}\bigr)_{13} = B_{\mu}$,
and $( \hat{Q}^{-1} \partial_{\mu}\hat{Q}\bigr)_{23} = C_{\mu}$,
which leads to the action
\bea
S_{\rm qnl\sigma m}&=& - \int {\rm d}^2x {\rm d}\tau
\, \Bigl\{ 2\, p_1^{\tau \tau}
A_{\tau}^2 + ( p_1^{\tau \tau} +p_3^{\tau \tau})(B_{\tau}^2
+C_{\tau}^2) \nonumber\\
& &{}+ \sum_{\alpha=x,y} \bigl[ 2\, p_1^{\alpha
\alpha} A_{\alpha}^2 + ( p_1^{\alpha  \alpha} +p_3^{\alpha
\alpha} )(B_{\alpha}^2+C_{\alpha}^2)\bigl] \Bigr\}
\eea
The dispersion relations for the three spin-waves can now be read
off:
\bea
\omega_{\scriptscriptstyle A}^2&=& 
\frac{p_1^{xx}}{p_1^{\tau \tau}}\, (k_{\scriptscriptstyle A})_x^2 +
\frac{p_1^{yy}}{p_1^{\tau \tau}} \,(k_{\scriptscriptstyle A})_y^2 
\,, \nonumber\\
\omega_{\scriptscriptstyle{B,C}}^2&=& 
\frac{(p_1^{xx}+p_3^{xx})}{(p_1^{\tau \tau}+p_3^{\tau \tau})}\,
(k_{\scriptscriptstyle{B,C}})_x^2 +
\frac{(p_1^{yy}+p_3^{yy})}{(p_1^{\tau \tau}+p_3^{\tau \tau})} \,
(k_{\scriptscriptstyle{B,C}})_y^2 \,.
\eea
Thus, the spin-waves have
different velocities which may depend on the 
direction of propagation, as was suggested by the hydrodynamical
theory \cite{halpsas}. It is the fact that the theory is not
Lorentz invariant, i.e.~that the coupling-constant matrix
$\hat{P}^{\mu \nu}$ for the
time components  is not proportional to that for the space components,
which allows the three spin-waves to have different velocities.
In our case, the two out-of-plane modes have the same spin-wave 
velocity, which differs from that of the in-plane mode. In the
case in which the coupling-constant matrix is isotropic,
i.e.~$\hat{P}^{xx}=\hat{P}^{yy}$, 
the spin-wave velocities become 
\be
c_{\scriptscriptstyle A}= \sqrt{ p_1^{xx}/p_1^{\tau \tau}} \label{Ca}
\ee
in the plane spanned by the classical ground state, and
\be
c_{\scriptscriptstyle{B,C}}=
\sqrt{ (p_1^{xx}+p_3^{xx})/(p_1^{\tau \tau}+
p_3^{\tau \tau})} \label{Cbc}
\ee
out of  this plane. 

Now we briefly consider the internal-space symmetries 
of the action which we have
obtained in Eq.~(\ref{act1}) or (\ref{act2}). This action is invariant
under global left O(3) rotations 
$\hat{Q} \rightarrow \hat{U}
\hat{Q}$, 
where $\hat{U}\, \epsilon\, O(3)$. This symmetry 
corresponds to the usual 
invariance under rotations of the basis vectors in spin space:
$n_c^a \rightarrow U^{ab} \,n_c^b$.
In addition, the action is invariant
under global right transformations 
$\hat{Q} \rightarrow \hat{Q}\hat{V}$
if $[\hat{P},\hat{V}]=0$.
The dimension of
the group to which $\hat{V}$ belongs depends on the values
of the three coupling constants contained in $\hat{P}$. 
In our case, in which two of the coupling constants are
equal, $ \hat{V}\, \epsilon\, O(2)$. 
This right transformation
$\hat{V}$ on $\hat{Q}$ corresponds to a mixing of the basis
vectors: $n_c^a \rightarrow \sum_b n^a_b \,V_{bc}$. Since 
$p_1 = p_2 \neq p_3$, the mixing occurs between the basis
vectors $\vec{n}_1$ and  $\vec{n}_2$.
The right O(2) invariance reflects the screw-axis-like  symmetry
of the spiral state, which is invariant under a rotation about the
axis $\vec{n}_3$ by an angle enclosed between a pair of spins,
followed by a lattice translation along the direction which connects
the two spins. Thus, our action for the quantum nonlinear $\sigma$ model
possesses an O(3)$\otimes$O(2) symmetry as in previous work for frustrated 
quantum spin systems \cite{azaria1,azaria0,azaria2}.

In addition to the SO(3) quantum nonlinear $\sigma$ model, 
we find in our microscopic
derivation the contribution 
(see Eqs.~(\ref{1stres}), (\ref{Sreal}))
\begin{equation}
S_{\rm lin}= \sum_{\alpha=x,y} \Bigl[\,\frac{1}{a^2}\,
	\frac{\partial}{\partial
	q^{\alpha}}\Phi^-_1(q)\Bigr|_{q=0} - \frac{J_{\rm H} 
	S^2 }{a} \, \sin (\delta Q_{\alpha}a)\, \Bigr] \int\! 
	{\rm d}^2x {\rm d}\tau\,
\,\vec{n}_1\cdot\partial_{\alpha}\vec{n}_2 \,,
\end{equation}
which is linear in the derivatives.
The action $S_{\rm lin}$ is of course invariant under rotations
of the basis vectors, $n_c^a\rightarrow U^{ab} \,n_c^b$. Under the 
transformations $n_c^a\rightarrow \sum_b n^a_b \, V_{bc} $, however, 
$S_{\rm lin}$ is invariant only if $\det{V}= +1$. Transformations 
with $\det{V}=-1$ change our right-handed set of basis vectors
into a left-handed one. This means that $S_{\rm lin}$ is not
invariant under a change of helicity of the spiral. 

Since the linear term is not positive definite, the weight of some 
field configurations in the path integral will tend to infinity,
and hence this term leads to instabilities. To recover a stable
ground state, we must ensure that the action is at a minimum. Thus,
we impose the condition that the prefactors of the linear 
contributions must vanish:
\begin{equation}
\frac{1}{a} \,\frac{\partial}{\partial
q^{\alpha}}\Phi^-_1(q)\Bigr|_{q=0}  \; {\stackrel {!}{=}}\;
 J_{\rm H} S^2\, \sin (\delta Q_{\alpha}a) \,.
\label{pitchcond}
\end{equation}
This condition yields two equations (one for $\alpha = x$ and
one for $\alpha = y$) 
for the two spiral pitch
parameters $\delta Q_x$ and $\delta Q_y$, whose values are then
determined as a function  of the microscopic parameters and 
the doping concentration. 
Thus, the wave vector of the spiral background is
self-consistently determined from the stability argument.
In order to check the consistency of the stability condition,
Eq.~(\ref{pitchcond}), 
we consider again the special cases of the spiral wave vector 
mentioned above. For a spiral wave vector in the (1,0) direction,
the condition in Eq.~(\ref{pitchcond}) is
trivially fulfilled  for $\alpha = y$ because  both sides  vanish 
(the symmetry relation, Eq.~(\ref{g0sym2}), implies that 
$\frac{\partial}{\partial q^{y}}\Phi^-_1(q)\bigr|_{q=0}=0$ for
this choice of $\vec{k}_s$). If the spiral 
wave vector lies along the (0,1) direction, the terms on both sides of
Eq.~(\ref{pitchcond}) vanish for $\alpha =x$. All other choices
for  $\vec{k}_s$ lead to nonvanishing terms on both sides of
Eq.~(\ref{pitchcond}) for $\alpha =x,y$.
 
Finally, our microscopic derivation gives rise to the following term 
which is third order in the fields and second order in the derivatives:
\begin{equation}
S_{\rm geom}=-\hat{\chi}^{\mu \nu}\,\int\! {\rm d}^2x {\rm d}\tau
\,\vec{n}_3\,\cdot
\, \bigl(\partial_{\mu} 
\vec{n}_3 \times \partial_{\nu}\vec{n}_3 \bigr)\,.
      \label{Sgeom}
\end{equation}
 From the definition $\hat{\chi}^{\mu \nu} = \frac{1}{2a^2}
\left[\bar{\chi}_1+\bar{\chi}_2 \right]^{-1}
\bigl[\chi_6^{\mu} +\bar{\chi}_7^{\mu}+\chi_8^{\mu}\bigr]
\chi_4^{\nu} $ (see Eq.~(\ref{hatchi})), and 
using Eq.~(\ref{suszcomp}), it is seen that $S_{\rm geom}$ is nonzero 
only for $\mu = \tau$ and  $\nu = x,y$. For $\delta \vec{Q} =0$, 
$\hat{\chi}^{\mu \nu}$ can be shown to  be zero, so that $S_{\rm geom}$ 
vanishes. Performing the first-order derivatives in 
Eqs.~(\ref{chi4}), (\ref{chi6})-(\ref{chi8}) and looking at 
Eq.~(\ref{barchis}), one finds
that $\hat{\chi}^{\tau \alpha}$ is purely imaginary.

As is well-known, a term having the same structure as $S_{\rm geom}$ appears 
in the continuum theory of one-dimensional quantum Heisenberg antiferromagnets 
with an order parameter in $S_2$. The Pontryagin index 
\be
{\cal Q} =\frac{1}{4\pi} \int {\rm d}x {\rm d}\tau\,
 \vec{n}\,\cdot
\,\bigr(\partial_{x}\vec{n} \times  \partial_{\tau} \vec{n} \bigl)
\ee
describes the winding number of the mapping $\pi_2(S_2)= {\Bbb Z}$. The 
long-wavelength action of the spin-$S$ antiferromagnetic Heisenberg chain
contains the term $S_{\vec{\scriptstyle A}}^{AF}=i 2\pi S {\cal Q}$,
which has led Haldane \cite{hald83} to conjecture that integer spin chains
posses a gap in the excitation spectrum in contrast to half-integer ones.
In a two-dimensional square lattice, however, the topological term is 
cancelled, since the summation of all rows of spins yields
$\lim_{a \rightarrow 0} \sum_q (-1)^q\, i2\pi S\, {\cal Q}(q)=
\int {\rm d}y \, i\pi S\, \partial_y {\cal Q}(y)=0$ \cite{topolaf}. A similar 
situation arises in the triangular lattice, where the corresponding 
topological term does not contribute to the dynamics \cite{wintel}.  
Using  the  symmetry properties of the coefficient $\hat{\chi}^{\mu \nu}$, 
we can write our geometric term as
\bea
S_{\rm geom}&=& {\textstyle \sum_{\alpha}}\, \hat{\chi}^{\tau \alpha} 
\int {\rm d}x\, {\rm d}y\, {\rm d}\tau \,\vec{n}_3\,\cdot
\, \bigr(\partial_{\alpha} \vec{n}_3 \times \partial_{\tau}\vec{n}_3\bigl)
\nonumber \\
&=& 4\pi\, \hat{\chi}^{\tau x} \int {\rm d}y\, {\cal Q}(y)
+  4\pi\, \hat{\chi}^{\tau y} \int {\rm d}x\, {\cal Q}(x)\,.
\eea
While ${\cal Q}$ is an integer, the coefficients in front of the topological 
terms $4\pi\hat{\chi}^{\tau \alpha}$ are not necessarily integer multiples 
of $\pi$ as in the undoped models, but are in general functions of the 
microscopic parameters and the doping. Therefore, the obtained field theory
contains two $\theta$-vacuum terms \cite{theta} (one for each spatial 
direction) with continuously varying parameters, in contrast to the until now 
known field theories obtained for quantum spin systems in two dimensions.

It should be noted that $S_{\rm geom}$ does not correspond to the 
topological invariant classifying the mapping $\pi_3(SO(3))= {\Bbb Z}$, which 
is third order in derivatives and therefore is not contained in our expansion. 
It is given by \cite{jackiw}
\begin{equation}
\tilde{\cal Q}  =1/(24\pi^2) \int {\rm d}^2x {\rm d}\tau
\, 
\epsilon^{ \lambda \mu \nu} \,
\mbox{Tr}\, \bigl[(\hat{Q}^T \partial_{\lambda}\hat{Q})
(\hat{Q}^T \partial_{\mu}\hat{Q})
(\hat{Q}^T \partial_{\nu} \hat{Q})\bigr]\,, \label{so3top}
\end{equation}
If we write $S_{\rm geom}$ in terms of the SO(3) matrix $\hat{Q}$, it becomes
\be
S_{\rm geom}= -f^{\mu \nu}\,\int\! {\rm d}^2x {\rm d}\tau
\,\mbox{Tr}\, \bigl[ \hat{C} \bigl(\hat{Q}^T
\partial_{\mu}\hat{Q}\bigr) \bigl(\hat{Q}^T 
\partial_{\nu} \hat{Q}\bigr)\bigr]\,,
\ee
where $C_{12}=-C_{21}=1$  and $C_{ij}=0$ otherwise, and we have introduced 
$f^{\mu \nu}=\frac{1}{2} (\hat{\chi}^{\mu \nu}-
\hat{\chi}^{\nu  \mu})$ 
which is completely antisymmetric in  the momentum-space
indices $\mu$ and $\nu$. A study up to ${\cal O} (a^3)$ is left for the future.

Since the variation of $S_{\rm geom}$ under an infinitesimal variation of 
the field vanishes, its value can only change in discrete steps and thus it 
is quantized. Field configurations falling in different homotopy sectors 
cannot be continuously deformed into one another. Non-trivial field 
configurations correspond in this case to instanton-like configurations for the
plane of the spiral. Since until now the few studies carried out for the 
$\theta$-term considered only the one-dimensional case \cite{theta}, such
that an extension to (2+1) dimensions is not straightforward, we restrict 
ourselves for the following discussion to the sector in which 
$S_{\rm geom}=0$, such that the results obtained previously in 
renormalization group analysis can be applied.

Renormalization group analyses of a $(2+1)$-dimensional SO(3) quantum 
nonlinear $\sigma$ model in the form which we have derived in 
Eq.~(\ref{diagact}) have been performed by Azaria {\em et al.} 
\cite{azaria1} and by Apel {\em et al.} \cite{apel}. 
Although previous microscopic derivations of the SO(3) quantum
nonlinear $\sigma$ model from frustrated Heisenberg models resulted 
in an action with $p_1^{\tau \tau}=p_3^{\tau \tau}$ and 
$p_3^{xx}=p_3^{yy}=0$ \cite{domb,antimotri,apel,ADM93}, 
the renormalization group 
calculations were carried out for the model which we obtained in 
Eq.~(\ref{diagact}), because the aforementioned conditions are not stable 
under renormalization. Azaria {\em et al.} \cite{azaria1} considered the 
case in which the coupling constant matrix is isotropic in lattice space, 
i.e.~$\hat{P}^{xx}=\hat{P}^{yy}$, so that the action contains four coupling 
constants, which correspond to
$p_1^{\tau \tau}$, $p_3^{\tau \tau}$, $p_1^{xx}$, $p_3^{xx}$
in our case. Instead of using these four coupling constants,
it is possible to work with two independent spin stiffnesses,
$\rho_{\scriptscriptstyle A} \propto p_1^{xx}$ and
$\rho_{\scriptscriptstyle B} \propto (p_1^{xx} +p_3^{xx})$,
and two uniform spin susceptibilities, 
$\chi_{\scriptscriptstyle A} \propto p_1^{\tau \tau}$ 
and $\chi_{\scriptscriptstyle B} \propto (p_1^{\tau \tau} +p_3^{\tau
\tau})$. (The spin-wave velocities are then given by
$c_{\scriptscriptstyle A}=\sqrt{\rho_{\scriptscriptstyle A}/
\chi_{\scriptscriptstyle A}}$  and 
$c_{\scriptscriptstyle B}=\sqrt{\rho_{\scriptscriptstyle B}/
\chi_{\scriptscriptstyle B}}$.)
In contrast to the case of the O(3) quantum nonlinear $\sigma$ model 
describing the long-wavelength properties of collinear spin configurations, 
here the spin-wave velocities $c_{\scriptscriptstyle A}$ and 
$c_{\scriptscriptstyle B}$ already renormalize at one-loop order. In fact, 
the spin-wave velocity in the O(3) quantum nonlinear $\sigma$ model at 
zero temperature is not renormalized at one-loop order \cite{chn} as a 
consequence of the Lorentz invariance of the theory. Such an invariance is 
absent in the present theory, as already pointed out before. 

The one-loop renormalization group equations for the parameters 
$c_{\scriptscriptstyle A}$, $c_{\scriptscriptstyle B}$,
$\rho_{\scriptscriptstyle A}$, $\rho_{\scriptscriptstyle B}$, 
$\beta$ admit a nontrivial fixed point for $T=0$, 
$c_{\scriptscriptstyle A}^{\ast}=c_{\scriptscriptstyle B}^{\ast}$ and
$\rho_{\scriptscriptstyle A}^{\ast}=\rho_{\scriptscriptstyle B}^{\ast}$.
At this point, the theory is O(3)$\otimes$O(3) $\sim$ O(4) symmetric
and Lorentz invariant. As in the case of the classical
O(3)$\otimes$O(2) nonlinear $\sigma$ model in $(2+\epsilon)$
dimensions \cite{azaria0,azaria2} (which contains two couplings in the 
isotropic case), the symmetry is dynamically enlarged at the fixed point.
Connected to this fixed point in the five-dimensional parameter space is
a critical surface separating an ordered phase, in our case a spiral
one, from a phase disordered by quantum fluctuations. The scaling 
properties of this quantum transition have been described by Chubukov 
{\em et al.} \cite{chub}. The present work provides a way to determine 
quantitatively how a variation of doping will either drive the system from 
the helically ordered to the gapped spin-liquid phase, or in case the 
spin-wave velocities vanish, an instability to another state occurs.
\section{Conclusion}
In this work, the long wavelength, low-energy sector of a spiral spin 
configuration in the microscopic spin-fermion model has been mapped onto an 
effective field theory. This mapping has been accomplished by a systematic 
gradient expansion of the spin action which was 
obtained by exactly integrating out 
the fermions from a coherent-state path-integral representation. The 
magnetically ordered spiral state completely breaks the O(3) rotation 
symmetry in spin space. This leads to an SO(3) order parameter. We have 
shown how the constituent fields of the order parameter are obtained from 
the physical spins, such that the ambiguity in describing the physical  
spin on $S_2$ with elements in SO(3) is avoided. The low-lying modes 
fluctuating around the classical spiral ground state were parametrized
by a decomposition of the spin fields into a helical and a uniform component.

The gradient expansion of the fermionic determinant leads to an infinite 
series that can be summed to all orders of the microscopic coupling constants 
by a combinatorial method which exploits the constraint on the order parameter. 
We have carried out the gradient expansion up to second order, such that the 
relevant terms in the same order as those generally proposed in 
phenomenological approaches are contained. Such an expansion leads to an 
effective action containing first- and second-order space-time derivatives of 
the order parameter. 

The first-order terms yield through a stability condition two equations which 
determine the two spiral pitch parameters as a function of the microscopic 
parameters and the doping. Thus, our theory is able to determine the wave 
vector of the spiral spin-background in a self-contained way.

The second-order terms, on the one hand, lead to an O(3)$\otimes$O(2)
symmetric quantum nonlinear $\sigma$ model describing the long-wavelength 
spectrum of three spin waves. The coefficient matrix of the $\sigma$ model, 
which is responsible for the Lorentz noninvariance of our theory, determines 
the spin-wave velocities and spin-wave stiffnesses as a function of the 
microscopic parameters and the doping. It should be emphasized that the 
continuum approximation only refers to the spin dynamics, so that our method 
can deal with arbitrary dispersion relations for the fermions. Since
the gradient expansion is nonperturbative in the coupling
constants of the microscopic Hamiltonian, its results
are also relevant for related models, like the Kondo lattice model ($J_H =0$)
and the $t$-$J$ model ($J_K \rightarrow \infty$). 

On the other hand, our continuum theory yields geometric terms of the same 
form as the one obtained for the one dimensional antiferromagnetic Heisenberg 
model in the continuum limit. In our case however, the corresponding 
coefficient varies continuously as a function of doping. Whether these terms 
may close the gap in the quantum disordered phase for a given value of the
coefficient (like at $\theta = \pi$ in nonlinear $\sigma$-models with a 
$\theta$ term in (1+1) dimensions), is a question left for further studies.  

Our results show that by dealing with a given but rather general
model for doped antiferromagnets, new terms appear that are in general not 
contained in phenomenological approaches, where only symmetry arguments are 
used. Furthermore, the influence of the doped charge carriers on the
critical behaviour of the incommensurate spin system is obtained explicitly.
The doping dependence is contained in the parameters of our SO(3) nonlinear 
$\sigma$ model through generalized fermionic susceptibilities, which in 
addition yield the new dispersion relations of the holes in the presence of 
spiral spin fields. The numerical evaluation of the generalized fermionic
susceptibilities is presently being carried out.

We would like to thank A.~Angelucci for useful discussions.
S.K.~gratefully acknowledges support by the Studienstiftung des
deutschen Volkes.

\begin{appendix}

\section{Generalized Fermionic Susceptibilities}
The quantities $\Phi_2^{\pm} \ldots \Phi_6^{\pm}$ which appear in 
the prefactor of the $(\partial_{\mu}\vec{n}_3 \cdot
\partial_{\nu}\vec{n}_3)$--term of the action Eq.~(\ref{2ndres})
describing the
spiral contribution to the fermion determinant are given by:
\begin{eqnarray}
\Phi^{\pm}_2(q)&=&\;\;\;\sum_k\, \mbox{Li}_2 (g^2 g_0(k\!-\!q)\, 
g_{0,\pm}(k\!-\!q))
\,, \label{Phi2}\\
\Phi^{\pm}_3(q)&=&\frac{1}{2}\sum_k\, \mbox{Li}_0 (g^2 g_0(k)\,
g_{0,\pm}(k))\, \,\frac{g_{0,\pm}(k\!-\!q)}{g_{0,\pm}(k)}\,,\\
(\Phi^{\pm}_4)^{\mu \nu}&=& \frac{1}{2}\sum_k\, \mbox{Li}_0 (g^2
g_0(k)\,g_{0,\pm}(k))\, \, \Bigl[\frac{g_{0,\pm}(k)}{g_{0,\mp}(k)}
-1\Bigr]^{-1}\nonumber \\ 
& &\frac{ [\frac{\partial g_0(k-q)}{\partial
q^{\mu}}]|_{q=0} }{g_0(k)} 
\frac{[\frac{\partial g_{0,\pm}(k-q)}{\partial q^{\nu}}]|_{q=0}}
{g_{0,\pm}(k)}\,,\\
(\Phi^{\pm}_5)^{\mu \nu}&=&\;\;\;\sum_k\,
\mbox{Li}_1(g^2g_0(k)\,g_{0,\pm}(k))
\Bigl[\frac{g_{0,\pm}(k)}{g_{0,\mp}(k)}-1\Bigr]^{-1}
    \Biggl\{ \frac{\partial}{\partial q^{\mu}} \Bigl[
      \frac{ \frac{\partial g_0(k-q)}{\partial q^{\nu}}}
      {g_0(k\!-\!q)} \Bigr] \Bigl|_{q=0}\nonumber\\
& &+\Bigl[\frac{g_{0,\mp}(k)}{g_{0,\pm}(k)}-1\Bigr]^{-1}
 \frac{[\frac{\partial g_0(k-q)}{\partial q^{\mu}}]|_{q=0} }
     {g_0(k)} 
     \Biggl[ \frac{[\frac{\partial g_{0,\pm}(k-q)}
     {\partial q^{\nu}}]|_{q=0}}{g_{0,\pm}} 
-\frac{[\frac{\partial g_{0,\mp}(k-q)}{\partial q^{\nu}}]|_{q=0}}
     {g_{0,\mp}}\Biggr] \Biggr\}\,, \\
(\Phi^{\pm}_6)^{\mu \nu}&=&\frac{1}{8}\sum_k\, \frac{g^2 \,
[\frac{\partial g_0(k-q)} {\partial q^{\mu}} ]|_{q=0} \,
[\frac{\partial g_{0,\pm}(k-q)} {\partial q^{\nu}} ]|_{q=0} }
{[1 - g^2 g_0(k)g_{0,\pm}(k)]^2}\,, \label{Phi6}
\end{eqnarray}

The generalized fermionic susceptibilities appearing in the
ferromagnetic contribution to the fermion determinant (see
Eq.~(\ref{Sferro})) are defined as follows:
\begin{eqnarray}
\chi_1&=& \sum_k  g^2 \,\Bigl\{ \Bigl[ g^2 -\frac{1}{2}\, g_0^{-1}(k)\, 
\varrho_+(k) \Bigr] d^{-1}(k)  \nonumber\\
& &- \Bigl[ g^2 -\frac{1}{2}\, g_0^{-1}(k)\,\varrho_+(k) \Bigr]^2 
\Bigl[ g^2 - g_{0,+}^{-1}(k)\, g_{0,-}^{-1}(k)  \Bigr] 
\, d^{-2}(k)\nonumber\\
& &- \frac{1}{4}\; g_0^{-2}(k)\;\varrho_-^2(k)\; \Bigl[ g^2+ 
g_{0,+}^{-1}(k)\, g_{0,-}^{-1}(k)\Bigr]\, d^{-2}(k) \Bigr\}\,, \\
\chi_2&=&-\frac{\chi_1 }{2}\,, \\
\chi_3&=& -\sum_k  g^4\, \varrho_-^2(k)\,\Bigl[
g^2+g_{0}^{-2}(k)\Bigr]\, d^{-2}(k)\,, \\
\chi_4^{\mu}&=& \sum_k \frac{1}{4} \, g^3\,\varrho_-(k) \,
 \frac{\partial}{\partial q^{\mu}} \Bigl[
g_0^{-2}(k\!-\!q) \,g_{0,+}^{-1}(k\!-\!q)\, g_{0,-}^{-1}(k\!-\!q) 
\label{chi4}
\nonumber \\
& &-g^2\, g_0^{-1}(k\!-\!q)\,
\varrho_+(k\!-\!q)\Bigr] \Bigl|_{q=0}\;
d^{-2}(k)\,,\\
\chi_5^{\mu}&=& \sum_k \frac{1}{6} \,
g^3\,\varrho_-^2(k)\,\frac{\partial}{\partial q^{\mu}}\Bigl[
g_0^{-3}(k\!-\!q)\Bigr] \Bigl|_{q=0}\; d^{-2}(k)\,,\\
\label{chi5}
\chi_6^{\mu}&=& \sum_k \frac{1}{4} \,g^3 \,
\frac{\partial}{\partial q^{\mu}} \Bigl\{
-g^2 \Bigl[ \frac{1}{2}g_0^{-1}(k)\varrho_-^{2}(k\!-\!q)
+\varrho_-^2(k)\,g_0^{-1}(k\!-\!q)\Bigr] \nonumber \\
& &+g_{0}^{-2}(k)\,\varrho_-(k) \Bigl[ 
g_{0,+}^{-1}(k)\,g_{0,-}^{-1}(k\!-\!q) 
- g_{0,-}^{-1}(k)\, g_{0,+}^{-1}(k\!-\!q) \nonumber \\
& &+\varrho_-(k)\, g_0^{-1}(k\!-\!q)\Bigr]
 \Bigr\} \Bigl|_{q=0}\; d^{-2}(k)\,,\label{chi6} \\
\chi_7^{\mu}&=& \sum_k g^3\, \Bigl[ g^2 -\frac{1}{2}\,
g_0^{-1}(k)\, \varrho_+(k) \Bigr] ^2 \frac{\partial}{\partial
q^{\mu}} \Bigl[g_0^{-1}(k\!-\!q)\Bigr] \Bigl|_{q=0}\; d^{-2}(k)\,,
\label{chi7}\\
\chi_8^{\mu}&=& \sum_k  \frac{1}{4} \,g^3 \,
\frac{\partial}{\partial q^{\mu}} \Bigl[
g_0^{-2}(k)\, \varrho_-(k)\, \varrho_+(k\!-\!q)\,\varrho_-(k\!-\!q)
\nonumber \\
& &+ \,2 g^2\, \varrho_-^2(k)\,g_0^{-1}(k\!-\!q) 
- g^2\,g_0^{-1}(k)\, \varrho_-^{2}(k\!-\!q)\Bigr] \Bigl|_{q=0} 
\; d^{-2}(k)\,. \label{chi8}
\end{eqnarray}
In the expression Eq.~(\ref{L}),
which is obtained for the ferromagnetic field
$\vec{L}$ from the saddle-point solution, the 
following abbreviations are used:
\begin{eqnarray}
\zeta_1&=&\frac{\bar{\chi}_1\,(\bar{\chi}_1+2\bar{\chi}_2 + 
\chi_3)}{2\,(\bar{\chi}_1+\bar{\chi}_2)^2
(\bar{\chi}_1+\chi_3)}\,, \\
\zeta_2^{\mu}&=&\frac{\chi_4^{\mu}\,
[(\bar{\chi}_1-\bar{\chi}_2)(\bar{\chi}_1+\chi_3)+
4\bar{\chi}_1 \bar{\chi}_2]}{2\,(\bar{\chi}_1+\bar{\chi}_2)^2 
(\bar{\chi}_1+\chi_3)}\,, \\
\zeta_3^{\mu}&=&\frac{\chi_6^{\mu}}{2\,(\bar{\chi}_1+\bar{\chi}_2)}+
\frac{(\bar{\chi}_7^{\mu}+\chi_8^{\mu})\,\bar{\chi}_2\,
(\chi_3-\bar{\chi}_1)}
{2\,(\bar{\chi}_1+\bar{\chi}_2)^2 (\bar{\chi}_1+\chi_3)}\,, \\
\zeta_4^{\mu}&=&\frac{\bar{\chi}_7^{\mu}\,(\bar{\chi}_2^2-\bar{\chi}_1
\chi_3)}{(\bar{\chi}_1+\bar{\chi}_2)^2 (\bar{\chi}_1+\chi_3)}-
\frac{\chi_5^{\mu}}{2\,(\bar{\chi}_1+\chi_3)}\,.
\end{eqnarray}
\end{appendix}

\newpage
\par

\end{document}